\documentclass[aps,pre,amsmath,superscriptaddress,amssymb,twocolumn]{revtex4}

\usepackage[normalem]{ulem}
\usepackage{graphicx}
\usepackage{amsmath,amssymb}
\usepackage{natbib}
\usepackage{color, hyperref}

\definecolor{mygrey}{gray}{0.80}
\definecolor{darkblue}{rgb}{0.0,0.0,0.5}
\hypersetup{colorlinks,breaklinks,
            linkcolor=darkblue,urlcolor=darkblue,
            anchorcolor=darkblue,citecolor=darkblue}

\begin{document}

\title{Phase Diagram of Hard Tetrahedra}

\author{Amir Haji-Akbari}
\affiliation{Department of Chemical Engineering, University of Michigan, Ann Arbor, Michigan 48109, USA}

\author{Michael Engel}
\affiliation{Department of Chemical Engineering, University of Michigan, Ann Arbor, Michigan 48109, USA}

\author{Sharon C. Glotzer}
\email{sglotzer@umich.edu}
\affiliation{Department of Chemical Engineering \& Department of Materials Science and Engineering, University of Michigan, Ann Arbor, Michigan 48109, USA}

\date{\today}

\begin{abstract}
Advancements in the synthesis of faceted nanoparticles and colloids have spurred interest in the phase behavior of polyhedral shapes. Regular tetrahedra have attracted particular attention because they prefer local symmetries that are incompatible with periodicity. Two dense phases of regular tetrahedra have been reported recently. The densest known tetrahedron packing is achieved in a crystal of triangular bipyramids (dimers) with packing density $4000/4671\approx85.63\%$. In simulation a dodecagonal quasicrystal is observed; its approximant, with periodic tiling $(3.4.3^2\!.4)$, can be compressed to a packing fraction of $85.03\%$. Here, we show that the quasicrystal approximant is more stable than the dimer crystal for packing densities below $84\%$ using Monte Carlo computer simulations and free energy calculations. To carry out the free energy calculations, we use a variation of the Frenkel-Ladd method for anisotropic shapes and thermodynamic integration.  The enhanced stability of the approximant can be attributed to a network substructure, which maximizes the free volume (and hence the ‚Äòwiggle room‚Äô) available to the particles and facilitates correlated motion of particles, which further contributes to entropy and leads to diffusion for packing densities below 65\%. The existence of a solid-solid transition between structurally distinct phases not related by symmetry breaking -- the approximant and the dimer crystal-- is unusual for hard particle systems.
\end{abstract}

\maketitle

\section{Introduction} \label{section:intro}

The self-assembly of nanoparticles into ordered structures is governed by interaction and shape anisotropy~\cite{GlotzerSolomonNatureMatrial2007}. Anisotropic particles are capable of stabilizing complex phases by entropy alone. Such structures can have potentially interesting optical and electrical properties yet to be fully investigated~\cite{ElSayedAccChemRes2004, BurdaChemRev2005, MurphyPhysChemB2005, NieNatureNano2010, Shevchenko2011}. Among anisotropic particles, tetrahedra are promising for assembling unusual structures because of their simplicity as well as their lack of inversion symmetry. When arranged face-to-face, tetrahedra form configurations with five-fold or icosahedral symmetries that are incompatible with periodicity. This results in geometric frustration and renders the assembly of tetrahedra more challenging than assembling other shapes. Various types of nano-tetrahedra have recently been synthesized from noble metals~\cite{KimAngewChemIntEd2004, DemortiereJPhysChemB2008} and crystalline silicon~\cite{Berenschot2009, Barett2009}. Micron-size colloidal tetrahedra made of colloidal spheres have also been reported~\cite{ManoharanPineScience2003}.  In certain cases, these tetrahedra may be treated as hard particles.

Particles whose interactions are dominated by repulsion can be modeled to first approximation as hard particles. Since all permissible configurations of such systems are of identical energy, entropic effects govern their phase behavior.  Classic examples of entropy-driven phase transitions are the isotropic-to-nematic transition for hard thin rods~\cite{Onsager1949} and the crystallization of hard spheres into close-packed structures upon compression~\cite{Kirkwood1951}. Entropy drives these particles to order, because doing so will increase the number of configurations accessible to the system. In other words, the increase in macroscopic (visible) order is accompanied by an increase in microscopic disorder (the number of microstates)~\cite{FrenkelPhysicaA1999}. The origin of ordering can also be explained by considering the underlying thermodynamics of hard particle systems. In the limit of infinite pressure, the Gibbs free energy $G=PV-ST$ is dominated by the $PV$ term, which means that the densest packing will be ultimately stable at sufficiently high pressures. To date, all known maximally dense packings of hard shapes are ordered~\cite{BezdekArxiv2010}.

Although the phase behavior of hard spheres has been investigated extensively~\cite{Nulero2008}, many fewer studies have been done on other hard shapes~\cite{EppengaFrenkel1984, VeermanPhysRevA1990,VeermanFrenkel1992, VegaJCP1992, BolhuisFrenkel1997, CampAllen1997, JohnEscobedo2008, RaduSchilling2009, EscobedoNatureMaterials2011}. A key feature of the reported phase diagrams is the occurrence of symmetry-breaking phase transitions (first and second order) in which the symmetry group of the high-density phase is a subgroup of the symmetry group of the low-density phase (see for example the phase transitions in ~\cite{EscobedoNatureMaterials2011}). This means that the compression of the isotropic fluid results in an increase of structural complexity by breaking at least one symmetry per transformation. For instance, hard cubes form a cubatic liquid crystal before crystallizing into a simple cubic lattice. In both the liquid crystal and the cubic crystal the rotational symmetry is broken while the translational symmetry is only broken in the crystal and is present in the cubatic phase~\cite{JohnEscobedo2008}.

\begin{figure}
\includegraphics[width=\columnwidth]{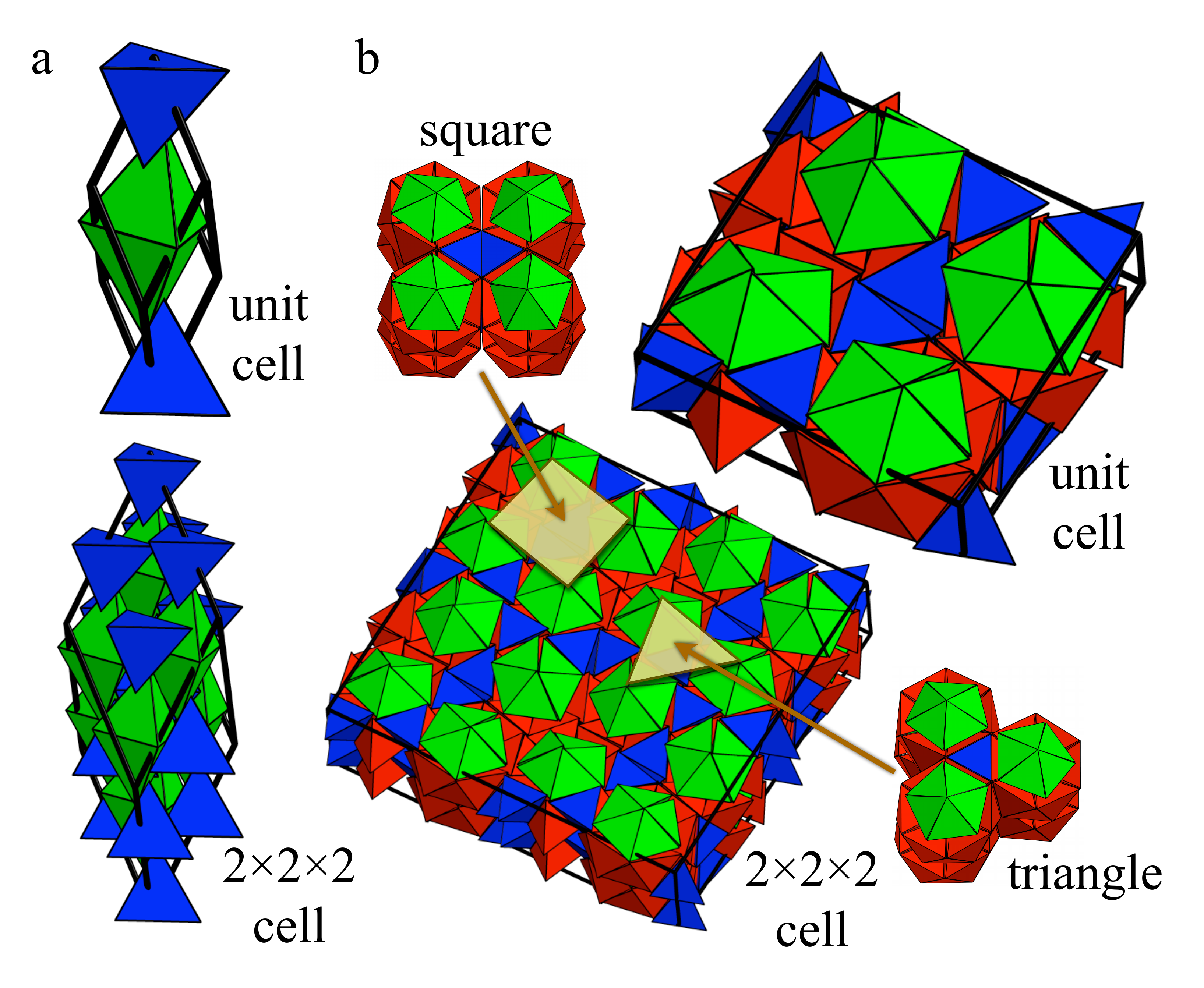}
\caption{\label{fig:structures}Dense packings of tetrahedra proposed by geometric construction and computer simulation. (a) The densest known tetrahedron packing~\cite{Chen2010} is a crystal with four particles per unit cell forming two triangular bipyramids (or `dimers')  shown in green and blue. The unit cell breaks the three-fold symmetry of the dimers. (b) In simulation, the particles form rings of twelve tetrahedra (red) capped by pentagonal dipyramids (green) together with interstitial tetrahedra (blue). The rings stack in logs, which arrange to form the vertices of a planar square-triangle tiling. Tilings observed in simulation are quasiperiodic, but a denser packing is obtained with the periodic $(3.4.3^2\!.4)$ Archimedean tiling, which is an approximant of the quasicrystal~\cite{HajiAkbariEtAl2009}.}
\end{figure}

The problem of assembling and packing hard tetrahedra has drawn significant attention over the last few years~\cite{ConwayTorquatoPNAS2006, ChenDSC2008, TorquatoJiaoNature2009, HajiAkbariEtAl2009, KallusArxiv2009, TorquatoJiaoarxiv, ChenEtAlarxiv2010, Kallus2010, Chen2010,  TJPRE2010, JaoshvilliPRL2010} and two competing phases have been reported in the high-density regime. The densest known packing of regular tetrahedra is a structurally simple double-triangular bipyramid crystal with packing density $\phi=4000/4671\approx85.63\%$ obtained from analytical construction and supported by numerical simulation~\cite{ChenEtAlarxiv2010, Chen2010}. It is obtained through optimizing an earlier monoclinic crystal discovered by Kallus \emph{et al}~\cite{KallusArxiv2009,Kallus2010}. We refer to this structure as the \emph{dimer crystal} throughout this work since the packing is characterized by pairs of tetrahedra incorporated into triangular bipyramids (Fig.~\ref{fig:structures}a).

Despite its stability in the limit of infinite pressure, simulations show that the dimer crystal does not form from the fluid except for systems of $16$ or fewer particles~\cite{Chen2010}. Instead, a dodecagonal quasicrystal spontaneously assembles at packing densities close to $50\%$ and above~\cite{HajiAkbariEtAl2009}. Structurally, the quasicrystal is significantly more complicated than the dimer phase; tetrahedra are arranged into rings that are further capped with pentagonal dipyramids (PDs). The rings and PDs are stacked in logs parallel to the ring axis, which in projection form the vertices of a planar tiling of squares and triangles (Fig.~\ref{fig:structures}b). Additional particles- referred to as intertitials-  appear in the space between the neighboring logs. It is noteworthy that the entire structure is a network of interpenetrating PDs spanning all particles in the system.  A periodic approximant of the quasicrystal, i.e. a crystal approximating the structure of the quasicrystal on a local level, with the $(3.4.3^2\!.4)$ Archimedean tiling and $82$ tetrahedra per unit cell compresses up to $\phi=85.03\%$, only slightly less dense than the dimer crystal~\cite{HajiAkbariEtAl2009}. In this paper we demonstrate that the approximant is more stable than the dimer crystal up to very high pressures and that the system prefers the dimer crystal thermodynamically only at packing densities exceeding $84\%$.

Here we carry out a detailed investigation of the phase behavior of hard tetrahedra from the fluid up to the densest packing. In contrast to previously studied systems of hard particles, the phase diagram of tetrahedra entails a non-symmetry-breaking solid-solid transition. We confirm its existence by Monte Carlo simulation and free energy calculations and discuss the origin of the transition. The present study complements previous works on hard tetrahedra which studied some aspects of the equation of state~\cite{GibbonsMolPhys1970, KolafaMolPhys1995, HajiAkbariEtAl2009} as well as dense packings~\cite{Kallus2010, Chen2010}, and extends those to provide a complete picture of the phase diagram. By comparing the results of self-assembly simulations to those obtained from free energy calculations, we assess the likelihood of various candidate phases to be observed both in simulations and in experiments of hard tetrahedra. 

The paper is organized as follows. The simulation methods as well as technical details of the free energy and free volume calculations are presented in Section~\ref{section:methods}. In Section~\ref{subsection:dimer_sym}, the thermodynamics of the dimer phase is reported. The thermodynamics of the quasicrystal and its approximant follows in Section~\ref{subsection:qcvsapp}. The results of free energy calculations are presented in Section~\ref{subsection:free_energy_results}. A computer experiment in which the dimer crystal spontaneously transforms into the quasicrystal at $\phi=50\%$ is reported in Section~\ref{subsection:dimer_to_qc}. The origin of the stability of the approximant over the dimer crystal at experimentally realizable densities is discussed in Section~\ref{subsection:origin_of_stability} and discussions and concluding remarks are provided in Section~\ref{section:conclusion}.

\section{Methods\label{section:methods}}

Simulations of $N$ hard, regular tetrahedra are carried out in the isochoric (NVT) ensemble and the isobaric (NPT) ensemble using a Monte Carlo algorithm. Forbidden overlaps of tetrahedra are determined using the separating axis theorem as explained in detail in Ref.~\cite{HajiAkbariEtAl2009}. $N$ particle trial moves are executed per Monte Carlo cycle. Each trial move can be a trial translation or a trial rotation chosen with equal probability. In the isobaric simulations, an additional box trial move is also performed where the size and shape of the simulation box are changed. The edge length of a tetrahedron, $\sigma$, is chosen as the unit length of the system. The effective pressure $P^*=P\sigma^3/k_BT$ is measured in dimensionless units. Maximum steps sizes are adjusted occasionally to allow for a target acceptance probability of $30\%$ and periodic boundary conditions are applied in all three dimensions.

\subsection{Equation of state}

Equations of state, $\phi(P^*)$, are calculated with isobaric simulations. Changes in the Gibbs free energy within a single phase are obtained via thermodynamic integration:
\begin{eqnarray}
\frac{G_2-G_1}{Nk_BT} &=& \frac{V_T}{\sigma^3}\int_{P_1^*}^{P_2^*}\frac{\text{d}p}{\phi(p)},\label{eq:free_energy_change_phase}
\end{eqnarray}
where $V_T=\sigma^3\sqrt{2}/12$ is the volume of a tetrahedron.

Simulations are carried out in the pressure range $50\le P^*\le4000$ for the dimer crystal ($4\times6\times6\times6 = 864$ tetrahedra), quasicrystal ($8,\!000$ tetrahedra) assembled from the fluid and compressed to a packing density up to $83.36\%$, and the approximant ($82\times2\times2\times3 = 984$ tetrahedra).

\subsection{Pressure estimation\label{subsection:pressure_estimation}}

The acceptance probability of trial volume changes is an estimator of the pressure in Monte Carlo simulations~\cite{FrenkelSmitBook}. Consider a trial expansion that increases the volume from $V$ to $V+\Delta V$. To fulfill detailed balance, the acceptance probability of the volume change is given by the Boltzmann factor,
\begin{equation}
P_{\text{B}}=\exp\left\{-\frac{P^*\Delta V}{\sigma^3}+
N\ln\left(1+\frac{\Delta V}{V}\right)\right\}.
\end{equation}
On the other hand, a trial compression that decreases the volume from $V$ to $V-\Delta V$ is accepted if and only if no overlap is generated by the trial volume change. Let $P_{\text{NO}}$ be the probability to generate an overlap in the trial compression. For small $\Delta V$ and in equilibrium the probabilities are equal, $P_{\text{NO}}=P_{\text{B}}$, and we can solve for the pressure:
\begin{eqnarray}
{P^*} = \lim_{\Delta V\rightarrow 0}\left\langle\frac{N\sigma^3}{\Delta{V}}
\left[\ln\left(1+\frac{\Delta{V}}{V}\right)-
\frac12\ln p_{\text{NO}}\right]\right\rangle.
\end{eqnarray}
Here, $p_{\text{NO}}=P_{\text{NO}}^{2/N}$ is the probability of a single particle not having any overlap with any other particle after the trial compression that decreases the volume by $\Delta V$.

\subsection{Free energy calculations}

\subsubsection{Frenkel-Ladd method for anisotropic hard particles\label{subsubsection:app_dimer_coex}}

The free energy of a (quasi-)crystal is calculated using the Frenkel-Ladd method~\cite{FrenkelLaddJCP1984, FrenkelSmitBook} by transforming it reversibly into an Einstein crystal, which serves as a reference structure with known free energy. In the Einstein crystal, each particle is tethered to its average lattice position via harmonic springs. Although originally developed for spherical particles, this method can be extended to particles with rotational degrees of freedom, such as tetrahedra. Additional springs are needed to tether the orientations of the particles to their average orientations in the lattice. Alternative extensions of the Frenkel-Ladd method to systems of particles with rotational degrees of freedom can be found in the literature~\cite{DijkstraPRE2008}.

We describe the configuration of a tetrahedron by $(\textbf{r}, \textbf{q})$, with $\textbf{r}$ being its center of mass position and $\textbf{q}$ the unit quaternion describing its orientation. The potential energy of of the corresponding Einstein crystal can then be expressed as:
\begin{eqnarray}\label{eq:Hamiltonian_Ein}
\frac{U(\textbf{r}^N,\textbf{q}^N)}{k_BT} =
\sum_{i=1}^N\frac{||\textbf{r}_i-\textbf{r}_{i,0}||^2}{\sigma^2} +
c\sum_{i=1}^N||\textbf{q}_i-\textbf{q}_{i,0}||^2
\end{eqnarray}
where $\textbf{r}_{i,0}$ and $\textbf{q}_{i,0}$ are the reference position and the reference orientation of the $i$-th particle in the crystal. The constant $c$ allows us to adjust the relative strength of the rotational springs and does not affect the computed free energy differences. All the results in this study are obtained using a value of $c=1/2$; we tested that using other values of $c$ does not affect the outcome of the calculations.

Each system is transformed to the Einstein crystal along a reversible path parameterized by $\gamma\in[0,\gamma_{\text{max}}]$ using the isochoric-isothermal (NVT) ensemble and the Hamiltonian
\begin{eqnarray}\label{eq:path}
\mathcal{H}(\textbf{r}^N,\textbf{q}^N;\gamma) =
\mathcal{H}_{\text{hard}}(\textbf{r}^N,\textbf{q}^N) +
\gamma U(\textbf{r}^N,\textbf{q}^N).
\end{eqnarray}
The hard particle system with Hamiltonian $\mathcal{H}_{\text{hard}}$ corresponds to $\gamma=0$, while in the limit $\gamma\rightarrow\infty$ the Einstein crystal is obtained. In practice, we can stop at a sufficiently large value of $\gamma_{\text{max}}$ when the springs are strong enough to suppress any particle collisions. The Helmholtz free energy difference $\Delta{A}=A_{\text{Ein}}-A_{\text{hard}}$ between the reference Einstein crystal and the hard particle system is given by:
\begin{eqnarray}\label{eq:path_integral}
\Delta{A} =
\int_{0}^{\gamma_{\text{max}}}\left\langle\frac{\partial\mathcal{H}(\gamma)}
{\partial\gamma}\right\rangle_{\gamma}\text{d}\gamma
= \int_0^{\gamma_{\text{max}}}\langle U\rangle_{\gamma}\text{d}\gamma
\end{eqnarray}
Note that the Frenkel-Ladd method can only be used if there is no translational or rotational diffusion in the system; otherwise the ensemble average $\langle{U}\rangle_{\gamma}$ will not be well-defined for small values of $\gamma$.

In our simulations,  the system is held for $2\times10^5$ Monte Carlo cycles at each $\gamma$ value during which $\langle U\rangle_{\gamma}$ is evaluated. The integral in equation (\ref{eq:path_integral}) is then computed numerically. This allows us to determine the Gibbs free energy $G=A+PV$ of the dimer (D) and the approximant (A) in the range $250\le P^*\le600$ where no configurational rearrangements are observed. The free energy difference $\Delta{G}=G_D-G_A$ is extrapolated to pressures outside this range using thermodynamic integration in addition to the Frenkel-Ladd method~\cite{EngelPRL2011}:
\begin{eqnarray}
\frac{\Delta{G}(P^*)}{Nk_BT} &=& \frac{\Delta{G}(P^*_0)}{Nk_BT}+
\frac{V_T}{\sigma^3}\int_{P^*_0}^{P^*}\left[\frac{1}{\phi_{D}(p)}-
\frac{1}{\phi_{A}(p)}\right]\text{d}p\notag\\
&& \label{eq:EOS_thermo_integration1}
\end{eqnarray}

\subsubsection{Fluid-solid transition\label{subsubsection:app_fluid_coex}}

We determine the melting pressure $P^*_M$ by calculating the absolute free energies of the solid and fluid. For sufficiently large values of $\gamma$, the Helmholtz free energy of the Einstein crystal is given by~\cite{FrenkelSmitBook}:
\begin{eqnarray}\label{eq:A_Einstein}
\frac{A_{\text{Ein}}}{Nk_BT} &=&
-\frac32\frac{N-1}{N}\ln\frac{\pi}{\gamma}
-\frac32\ln\frac{\pi}{c\gamma}-\ln N_{\text{sym}}\notag\\
&&+3\frac{N-1}{N}\ln\frac{\Lambda}{\sigma},
\end{eqnarray}
where $\Lambda=h/(2\pi{mk_BT})^{1/2}$ is the de Broglie wavelength. 
$N_{\text{sym}}$ is the number of quaternions corresponding to orientations that are symmetry-equivalent, which is twice the order of the rotation group of the particle. The factor $2$ arises from the fact that quaternions are inherently degenerate in describing the orientation i.e. $\textbf{q}$ and $-\textbf{q}$ correspond to the same rotation matrix. For a non-symmetric particle, the rotation group will have one element (identity) only and $N_{\text{sym}}=2$. Here, for tetrahedra, the rotation group has twelve elements, so $N_{\text{sym}}=24$. The first and the second terms are configurational contributions resulting from the translational and rotational springs. The last term corresponds to momentum contributions due to translational degrees of freedom. Momentum contributions due to rotational degrees of freedom are identical for the fluid and the solid and are therefore not included here.

The Gibbs free energy of an ideal gas, which approximates a real gas in the limit of infinite dilution, is 
\begin{eqnarray}\label{eq:ideal_G}
\frac{G_{\text{id}}}{Nk_BT}=\ln\frac{P^*}{2\pi^2}+
\frac{\ln(2\pi{N})}{2N}+3\ln\frac{\Lambda}{\sigma}.
\end{eqnarray}
The free energy of the fluid phase is then obtained from thermodynamic integration~\cite{RomanoJChemPhys2010}:
\begin{eqnarray}\label{eq:fluid_G}
\frac{G_{\text{fluid}}(P^*)}{Nk_BT}=\frac{G_{\text{id}}(P^*)}{Nk_BT}+
\int_0^{P^*}\left[\frac{V_T/\sigma^3}{\phi(p)}-\frac{1}{p}\right]\text{d}p.
\end{eqnarray}
We calculate $G_{\text{fluid}}(P^*)$ using the equation of state for a system
of $N=4,\!096$ tetrahedra for $0.01\le{P}^*\le60$.

\subsubsection{Finite size effects}

To ensure the system sizes we use are free of finite size effects, we calculate the Gibbs free energy difference between the dimer crystal and the approximant $\Delta{G}=G_D-G_A$, using equation (\ref{eq:A_Einstein}):
\begin{eqnarray}
\frac{\Delta{G}(N)-\Delta{G}}{Nk_BT} = \frac32\left[\frac{1}{N_{D}}-
\frac{1}{N_{A}}\right]\ln\frac{\pi\sigma^2}{\gamma_{\text{max}}\Lambda^2}
\end{eqnarray}
For the particle numbers used in the free energy calculations, $N_{D}=864$, $N_{A}=984$, $\gamma_{\max}=4\times10^6$ and $\sigma/\Lambda=\sqrt{2\pi}$, the error in $\Delta{G}$ is on the order of $10^{-3}k_BT$, which is negligible for the present purposes.

\subsection{Free volume calculations\label{subsection:free_volume_calc}}

The free volume of a hard sphere is the volume of the region of space in which the sphere can be moved continuously without overlapping with its neighbors while keeping all the other particles fixed~\cite{Hoover1968}. The definition generalizes to anisotropic particles with rotational degrees of freedom where free volume $v_{f}$ is now the volume of the largest subset of configurational space connected to the origin that can be accessed by a given particle while fixing the positions and orientations of all other particles~\cite{VegaMonsonMolPhys1992}:
\begin{eqnarray}\label{eq:free_volume_def}
v_{f} &=& \int I(\textbf{r},\textbf{q})
\text{d}^3\textbf{r}\text{d}^3\textbf{q}.
\end{eqnarray}
Here, $I(\textbf{r},\textbf{q})$ is the indicator function of motions $(\textbf{r},\textbf{q})$ consisting of a translation by $\textbf{r}$ and a rotation by $\textbf{q}$ and connected to the origin. $I$ is unity if the particle does not overlap with any other particle and zero otherwise. Due to the inherent periodicity of rotational motion, the free volume of an anisotropic particle has generally a more complicated topology compared to the free volume of a sphere. Here we calculate free volumes at high densities where the free volume is simply connected.

\subsubsection{Shooting method\label{subsubsection:shooting}}
We calculate the free volume of a particle using a method we call the \emph{shooting method}. Let $(\textbf{u},\textbf{v})$ correspond to a unit vector in the six dimensional configuration space and suppose that particle $i$ is `shot' in this direction until it hits another particle. The `shooting distance' is the smallest value of $\alpha$ for which the particle first overlaps with its neighbors if translated by $\alpha\textbf{u}$ and oriented according to the quaternion $(\textbf{q}_i+\alpha\textbf{v})/||\textbf{q}_i+\alpha\textbf{v}||$.

A lower bound for the free volume can be obtained by averaging over a sufficiently large number $N_s$ of shots with shot distances $\alpha_j$ along randomly chosen directions:
\begin{eqnarray}\label{Eq:shooting}
 v_{f} &\gtrapprox& \lim_{N_s\rightarrow\infty}\frac{1}{N_s}\sum_{j=1}^{N_s}
\frac{\pi^3}{6}\alpha_j^6.
\end{eqnarray}
Here $\pi^3/6$ is the volume of the six-dimensional unit sphere. Note that the periodic topology and the curvature of the six-dimensional configuration space are ignored, which is acceptable at high packing densities because $||\Delta\textbf{q}||\ll1$.

Eq. (\ref{Eq:shooting}) is a lower bound for concave free volumes, because shooting only allows access to the parts of the free volume connected to the origin by a straight line. Non-convex free volumes can arise from sliding collisions which, however, become increasingly rare at high packing densities. In fact, as we will show now for tetrahedra, the shooting method is accurate for high enough packing densities.

\subsubsection{Binning method}

To estimate the amount of error in the shooting method introduced by non-convexity, we using the alternative \emph{binning method} which corresponds to a Monte Carlo integration of the free volume. The configuration space of a given particle is partitioned into $N_{\text{bins}}$ small radial bins of volume $V^{\text{bin}}$. We perform $N_t$ random ghost trial moves per bin to average out the orientational degrees of freedom and determine the number $N^{\text{NO}}$ of trial moves not leading to an overlap. Free volume can then be estimated from:
\begin{eqnarray}
v_{f} \lessapprox
\frac{1}{N_t}\sum_{j=1}^{N_{\text{bins}}} V^{\text{bin}}_jN^{\text{NO}}_j.
\end{eqnarray}
Binning is much slower than shooting and might overestimate the free volume, if a trial move discovers an area of configuration space without overlap, but not connected to the original particle position. We find that the average of the logarithms of the free volumes calculated from the shooting method and the binning method agree within a relative error of $10^{-2}$ for all densities $\phi\ge70\%$.

\subsubsection{Mean-field approximation\label{MeanFieldDefine}}

The distribution of free volumes is related to the entropy of a hard particle system in the mean-field approximation. If we assume that free volumes of neighboring particles are uncorrelated, then the partition function of the system is expressed as $Q_{\text{mf}}=\prod_{i=1}^Nv_{f,i}$ and the Helmholtz free energy as ${A}_{\text{mf}}/Nk_BT=-\langle\ln v_{f}\rangle$. The thermodynamically relevant quantity is therefore the mean-log average of free volumes:
\begin{eqnarray}
v_{f,\text{ML}} &:=& \exp\langle\ln v_{f}\rangle\label{eq:vfML}
\end{eqnarray}  
which will be used in the rest of this study instead of the simple average $\langle v_{f}\rangle$.

\section{Results}\label{section:results}

\subsection{Symmetrization of the dimer packing on decreasing pressure}
\label{subsection:dimer_sym}
\begin{figure}
\includegraphics[width=\columnwidth]{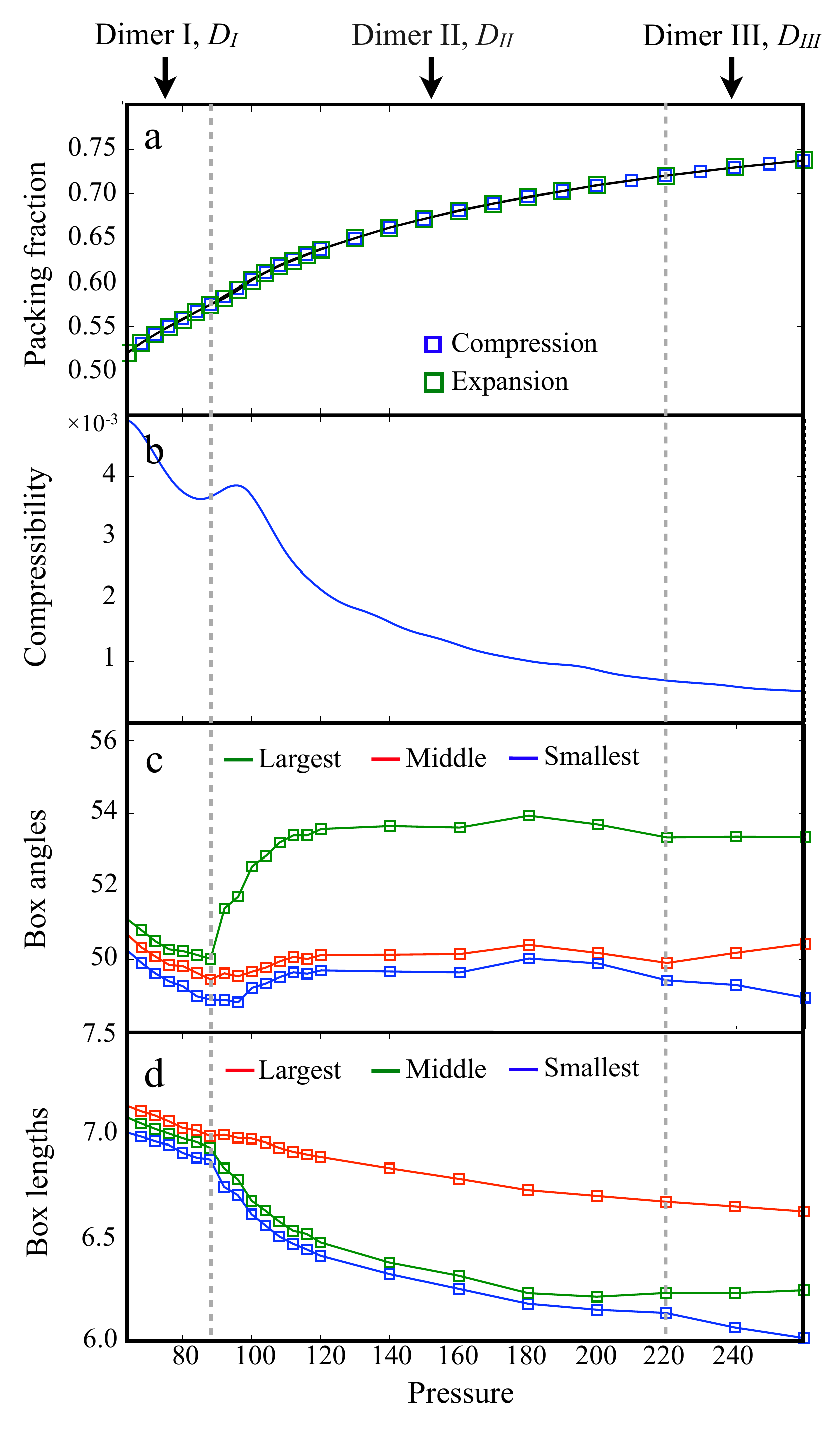}
\caption{\label{fig:dimer_phase}Symmetrization of the dimer crystal. (a) The equation of state shows no hysteresis between compression and expansion. (b) A peak is observed in the compressibility near pressure $P^*=90$ at a second order displacive phase transition. The (c) three box angles and (d) box lengths, obtained by sorting the angles and lengths and then averaging the sorted values, are plotted as a function of pressure. We observe two transitions, from triclinic ($\text{D}_{III}$) to monoclinic ($\text{D}_{II}$) to rhombohedral ($\text{D}_I$). The phase $\text{D}_{III}$ is thermodynamically stable, $\text{D}_{II}$ and $\text{D}_{I}$ are metastable.}
\end{figure}

We construct the dimer crystal analytically~\cite{Chen2010} and slowly expand it by reducing the pressure. The crystal remains stable during the simulation for pressures $P^*\ge60$ while at lower pressures it melts abruptly. No hysteresis is observed in the equation of state (Fig.~\ref{fig:dimer_phase}a), if the decompression is stopped before melting and the system is re-compressed. This suggests that the system remains at least in metastable equilibrium over this range of pressures and densities.

The compressibility $\kappa= (1/\phi)(\partial\phi/\partial P^*)$ (Fig.~\ref{fig:dimer_phase}b) reveals a complicated phase behavior, with an anomalous peak indicative of a second-order phase transition appearing at around $P^*=90$. We verify that this is a displacive phase transition; i.e. it only involves a lattice distortion and the particles in the lower density phase still remain in dimers. Analyzing the lengths of the vectors spanning the simulation box and the angles between them (Figs.~\ref{fig:dimer_phase}c,d) indicates that the transition takes place in two stages. While in the lower density phase $\text{D}_{I}$ ($P^*<90$) all lengths and angles are equivalent, they are completely split only in the phase $\text{D}_{III}$ ($P^*>220$). There is also an intermediate phase $\text{D}_{II}$ ($90<P^*<220$) in which only two of the lengths and angles are still degenerate. The symmetrization of the lattice therefore follows the sequence: triclinic ($\text{D}_{III}$) $\rightarrow$ monoclinic ($\text{D}_{II}$) $\rightarrow$ rhombohedral ($\text{D}_{I}$).

It is known that the three-fold symmetry of the dimers must be broken to achieve optimal bulk packing ~\cite{Kallus2010, Chen2010}, and we observe this in the sequence of transitions. We note that $\text{D}_{II}$ was initially reported by Kallus \emph{et al.} as a candidate for the densest packing of tetrahedra~\cite{Kallus2010}. Its maximum packing density is only $0.2\%$ lower than the maximum packing density of $\text{D}_{III}$, the structure predicted by Chen \emph{et al.}~\cite{Chen2010}. Note also that the integrated area under the peak is a measure of the difference in packing densities. This explains the missing peak in the compressibility for the transition $\text{D}_{III}\rightarrow\text{D}_{II}$. In contrast, the difference in maximum packing densities for the transition  $\text{D}_{II}\rightarrow\text{D}_{I}$ is much larger, and of the order of a few percent.

\subsection{Comparison of the quasicrystal and its $(3.4.3^2.4)$ approximant\label{subsection:qcvsapp}}

\begin{figure}
\includegraphics[width=\columnwidth]{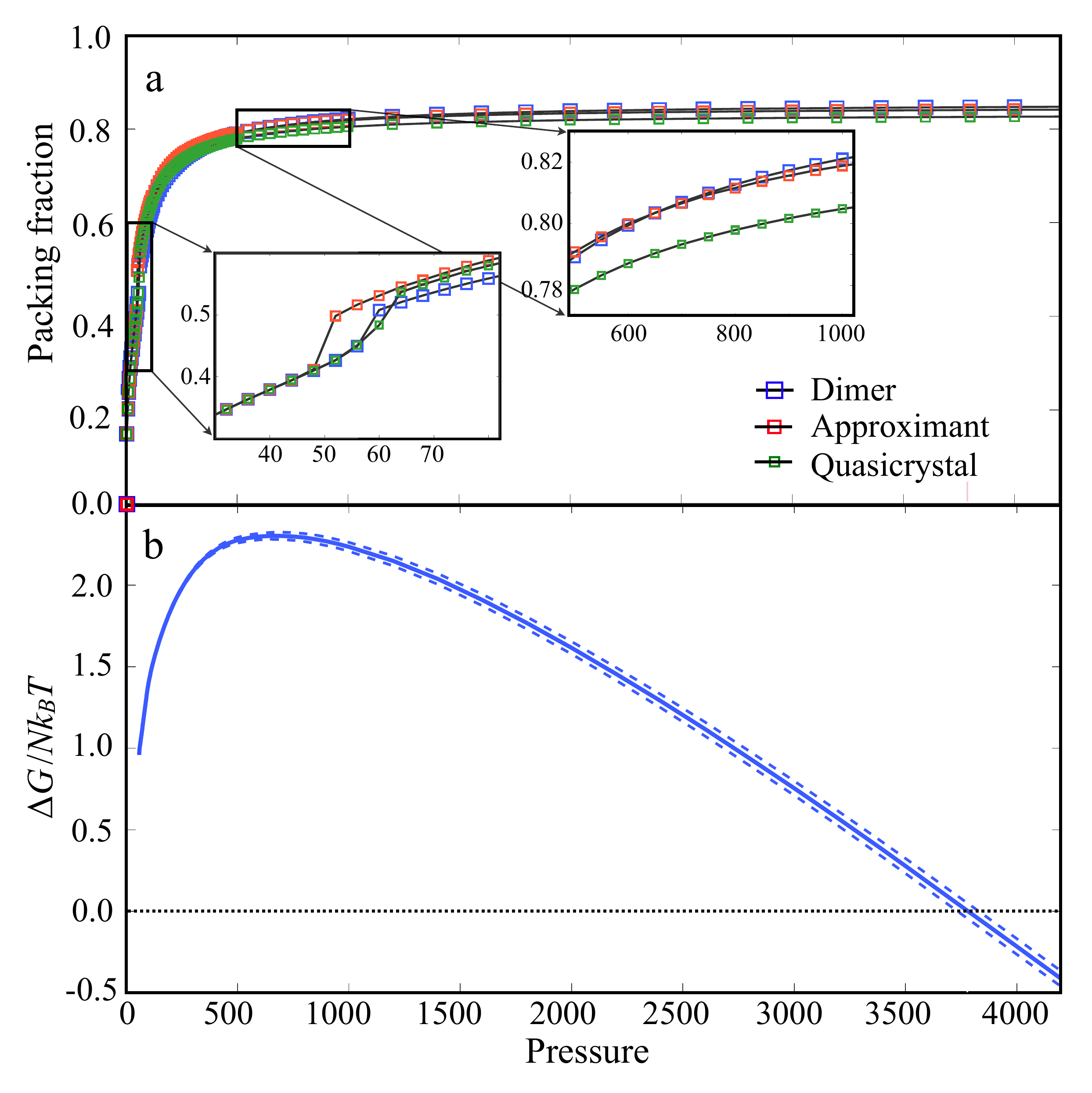}
\caption{Thermodynamic stability of the dimer crystal. (a) The equation of state for the dimer crystal, the approximant, and the quasicrystal shows that the dimer crystal is the densest packing for $P^*>700$. The approximant is always denser than the quasicrystal. Error bars are smaller than the size of the symbols. Insets show the equations of state in the melting region as well as near $P^*=700$ where the dimer crystal first becomes denser than the approximant. (b) The Gibbs free energy difference between the dimer crystal and the approximant $\Delta{G}/Nk_BT=(G_D-G_A)/Nk_BT$ calculated using thermodynamic integration and the Frenkel-Ladd method. The dimer crystal is stable only at very high pressures.\label{fig:EOS_free_energy}}
\end{figure}

The equations of state of the quasicrystal, the approximant, and the dimer crystal are presented in Fig.~\ref{fig:EOS_free_energy}a. We observe that the approximant is not only denser than the quasicrystal at all pressures above the melting transition, it also melts at lower pressure. These observations together with Eq.~(\ref{eq:free_energy_change_phase}) suggest that the quasicrystal is generally less stable than the approximant.

Further evidence for the stability of the approximant over the quasicrystal is obtained through constructing higher order approximants, i.e. approximants that have larger unit cells than the $(3.4.3^2\!.4)$ approximant, and comparing their equations of state with the quasicrystal and the approximant. For this purpose, we construct the second-order approximant with a unit cell containing $1,\!142$ tetrahedra using an inflation operation~\cite{StampfiiHelvPhysActa1986} and compute its equation of state near the transition region.

\begin{figure}
\includegraphics[width=0.9\columnwidth]{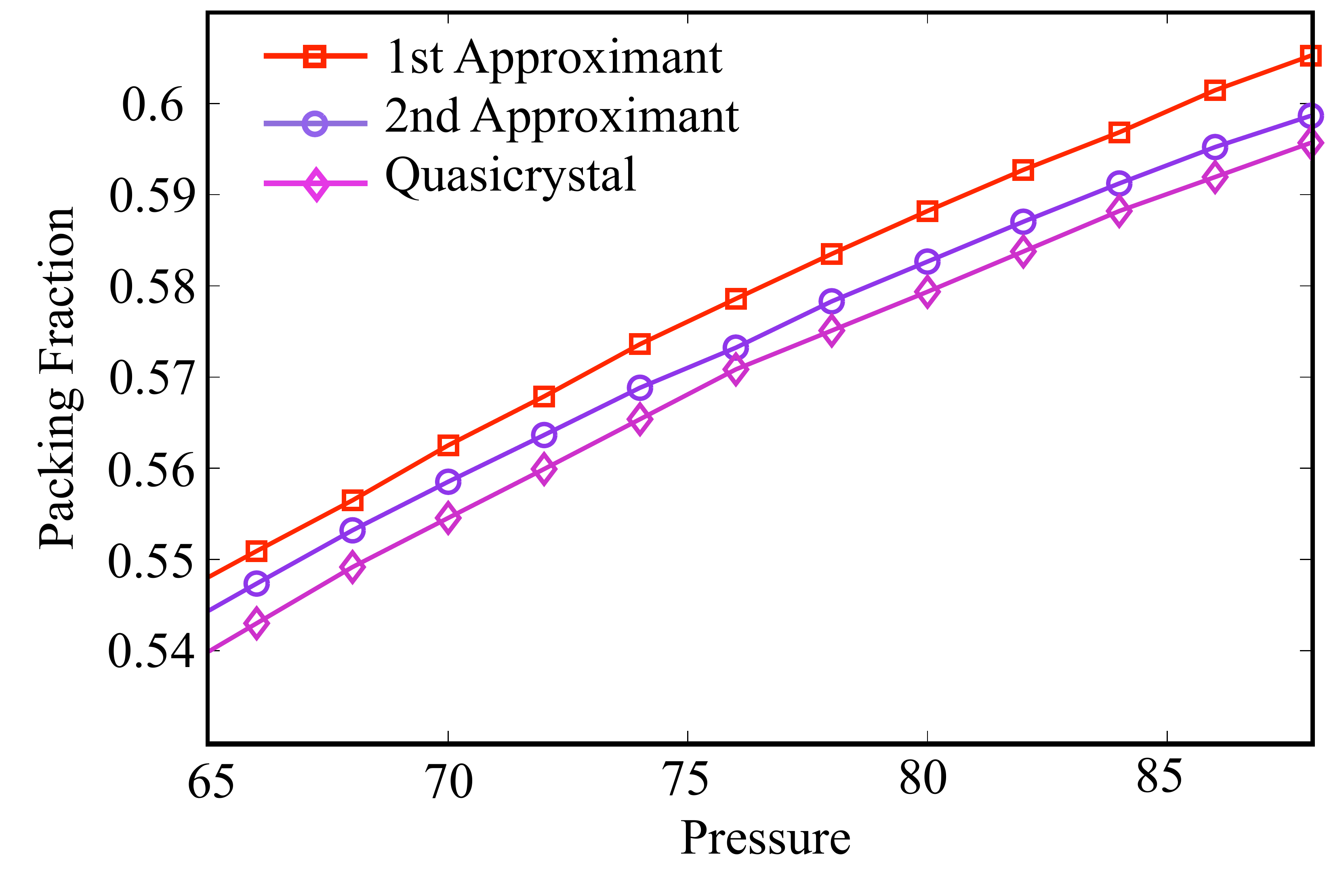}
\caption{Equations of state of the quasicrystal and the first and second approximants computed from NPT simulations. The second approximant (with a unit cell of $1,\!142$ particles) is less dense than the first approximant (with a $82$-particle unit cell). \label{fig:DifferentApproximantsEOS}}
\end{figure}

As observed in Fig.~\ref{fig:DifferentApproximantsEOS},  the second approximant is denser than the densest quasicrystal that formed in our simulations but less dense than the first approximant. Neither structure is expected to have a significant entropic advantage over others since tetrahedra experience similar local environments in all these structures. It is therefore safe to conclude that  the first approximant is more stable than the quasicrystal and the second approximant because of its higher density. Higher-order approximants can be constructed similarly using inflation symmetry; however, such approximants will have very large unit cells with tens of thousands of particles. Based on the observed trend, we expect higher-order approximants to become successively less dense but still denser than the quasicrystal.

The question of comparing the relative thermodynamic stability of quasicrystals and their approximants plagues nearly all reports of new quasicrystals in the literature.  The difficulty in obtaining perfect quasicrystals in experiments and simulations, along with the slow kinetics that would be involved in the transformation of even an imperfect quasicrystal to any of its approximants, confounds attempts to address quasicrystal stability.  In this spirit, we remark that the quasicrystal configuration used in this study is obtained in simulation and an ideal, perfect quasicrystal might be slightly denser. The structure of such an ideal quasicrystal, however, is unknown. A denser quasicrystal would shift the curve in Fig.~\ref{fig:EOS_free_energy}a slightly upwards, and hence make the quasicrystal thermodynamically more stable than the approximant in a narrow region close to melting. Based on all evidence however, we use the $(3.4.3^2\!.4)$ approximant as the most stable quasicrystal-like structure for free energy and free volume calculation purposes. 

\subsection{Relative thermodynamic stability\label{subsection:free_energy_results}}

\begin{figure}
\includegraphics[width=0.9\columnwidth]{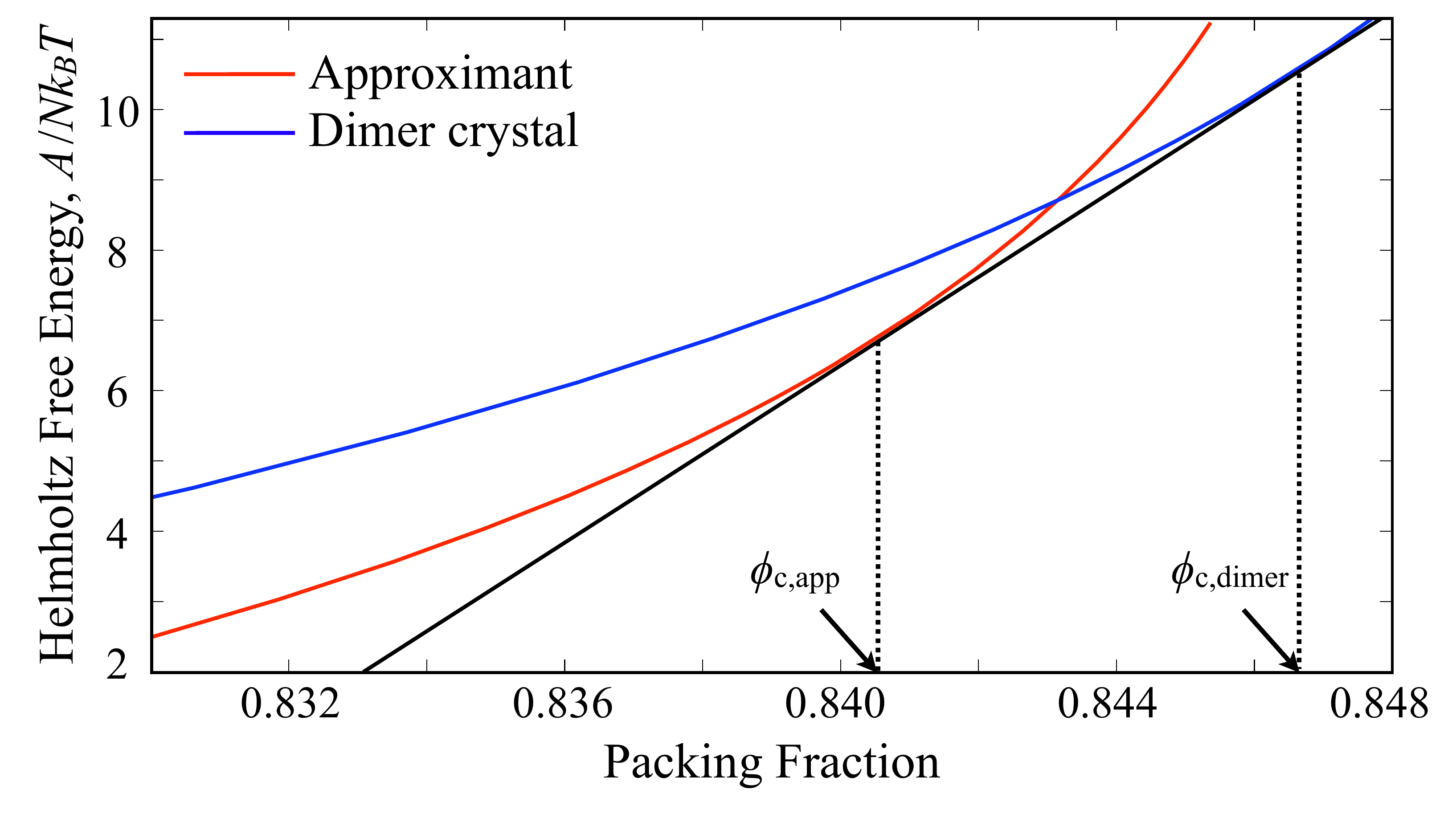}
\caption{The critical packing fractions for the approximant to dimer transition can be calculated via the common tangent construction from the Helmholtz free energies of the approximant (red) and the dimer crystal (blue). \label{fig:CommonTangentConstruction}}
\end{figure}

\begin{figure}
\includegraphics[width=0.9\columnwidth]{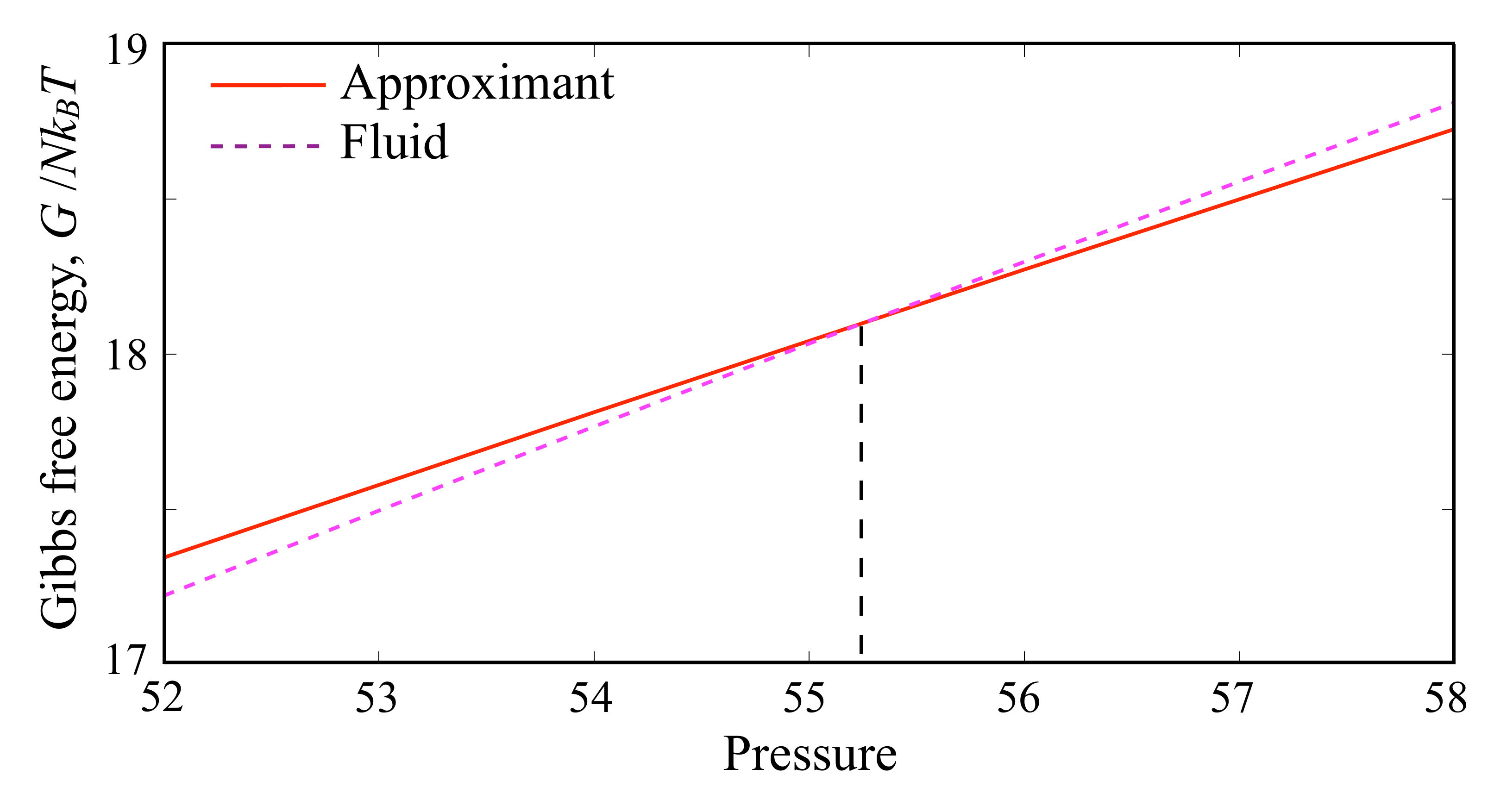}
\caption{Gibbs free energies of the approximant and the fluid close to the melting transition. The transition occurs at $P^*_M=55$. \label{fig:appFluidCoex}}
\end{figure}

The Gibbs free energy difference between the dimer and the approximant is calculated using the method described in Sec.~\ref{subsubsection:app_dimer_coex}. We find that the dimer crystal is stable only for pressures above $P^*_c = 3780\pm60$ (Fig.~\ref{fig:EOS_free_energy}b), while the approximant is favored below $P^*_c$. At the critical pressure, the approximant and the dimer crystal have packing densities of $\left(84.0\pm0.1\right)\%$ and $\left(84.6\pm0.1\right)\%$ respectively. The transition densities can be alternatively calculated from the Helmholtz energy using the common-tangent construction (Fig.~\ref{fig:CommonTangentConstruction}). $P^*_c$ is significantly higher than the melting pressure for the approximant, {Gray}$P^*_M=55\pm1$ (Fig.~\ref{fig:appFluidCoex}), which is determined using the approach described in Section~\ref{subsubsection:app_fluid_coex}.

It is noteworthy that the above calculations are based on the assumption that the dimer crystal of~\cite{ChenEtAlarxiv2010,Chen2010} is the densest possible arrangement of hard tetrahedra. Although we cannot rule out the possibility that an even denser arrangement of tetrahedra that is different from the approximant and the dimer crystal might exist, our observation that the dimer crystal is the densest structure that forms in simulations of $16$ tetrahedra and fewer~\cite{ChenEtAlarxiv2010,Chen2010} substantiates this assumption. The quasicrystal that we are using for comparison with the approximant has been assembled in simulations from the disordered fluid and therefore contains imperfections. We cannot rule out that a perfect quasicrystal might be thermodynamically more stable than the approximant at all pressures. If this were the case, then the transition between the approximant and the dimer crystal reported above would be substituted by a transition between the quasicrystal and the dimer crystal in the phase diagram. Therefore, while such a discovery could alter certain details of the phase transition, it will not eliminate the existence of a solid-solid phase transition reported in this work.

\subsection{Dimer-quasicrystal transformation\label{subsection:dimer_to_qc}}
To compare the relative thermodynamic stability of the dimer crystal and the quasicrystal in simulation, we set up a Monte Carlo simulation of a large dimer crystal with $2,\!916 (= 4\times9\times9\times9)$ tetrahedra in the isochoric ensemble. To facilitate the transformation, the box dimensions are occasionally distorted in a random direction with the constraint that the total volume remains unchanged (variable-shape ensemble,~\cite{VeermanPhysRevA1990}). This distortion allows the system to adjust to arbitrary lattice symmetries by relaxing shear stresses.

We choose a constant packing density of $\phi=50\%$, because at this density the quasicrystal is routinely observed to form spontaneously from the fluid. Structural changes are detected by counting the number of particles that form PDs and icosahedra using a shape-matching algorithm~\cite{KeysAnnRevCondMatPhys2011}; icosahdral motifs vanish when the quasicrystal forms~\cite{HajiAkbariEtAl2009}.  Additionally, the pressure is determined from the acceptance probability of trial volume changes as described in Section~\ref{subsection:pressure_estimation}~\cite{EppengaFrenkel1984, HarismiadisJChemPhys1996}.

\begin{figure}
\includegraphics[width=\columnwidth]{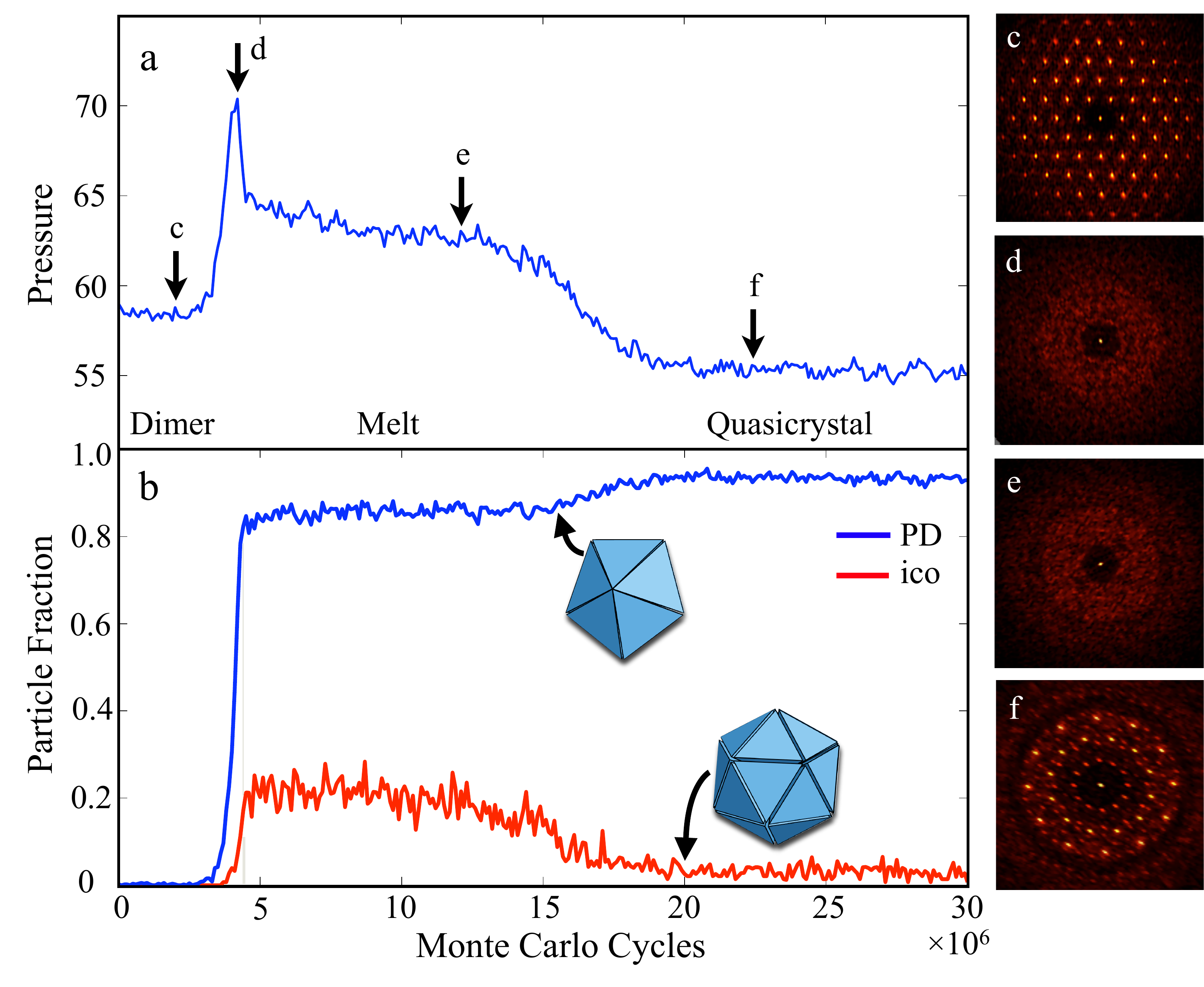}
\caption{Transformation of the dimer crystal to the dodecagonal quasicrystal in an isochoric simulation. (a) The pressure first spikes after $4$ million Monte Carlo cycles and then relaxes during the melting of the dimer crystal. Between $15$ and $20$ million Monte Carlo cycles, the quasicrystal forms from the melt. (b) The number of particles arranged in pentagonal dipyramids (PDs) or icosahedra (ico) increases rapidly during melting. In the quasicrystal essentially all particles form PDs while icosahedra disappear. Diffraction patterns confirm the transformation from the dimer crystal (c) to the melt (d,e) and then to the quasicrystal (f).\label{fig:DimerMelting}}
\end{figure}

The pressure shows a sharp spike after $4$ million Monte Carlo cycles accompanying the melting of the dimer crystal (Fig.~\ref{fig:DimerMelting}a). The spike quickly decreases to a plateau, which, after $15-20$ million Monte Carlo cycles, relaxes to its equilibrium value. PDs and icosahedra form as the preferred local configurations in the melt (Fig.~\ref{fig:DimerMelting}b). On the other hand, in the final solid structure, most particles are members of PDs and virtually no icosahedra remain. Diffraction images in Figs.~\ref{fig:DimerMelting}c-f show that the final solid structure is the dodecagonal quasicrystal. The fact that the quasicrystal forms in the simulation with the melt as an intermediate state confirms that both the quasicrystal and the melt are thermodynamically favored over the dimer crystal at the packing density $\phi=50\%$. 

\subsection{Origin of stability of the approximant\label{subsection:origin_of_stability}}
To investigate the superior stability of the quasicrystal approximant compared to the dimer crystal over such a wide range of densities, we investigate the significance of collective particle motions by comparing the free energy estimates obtained from the mean-field approximation introduced in Section~\ref{MeanFieldDefine} with the exact free energy differences. We also analyze the dynamics in the approximant by calculating the van Hove correlation function~\cite{vanHovePR1954} and visually inspecting the high-mobility particles \cite{KobPRL1997} in our simulations. 

\subsubsection{Free volumes}

\begin{figure}
\includegraphics[width=0.9\columnwidth]{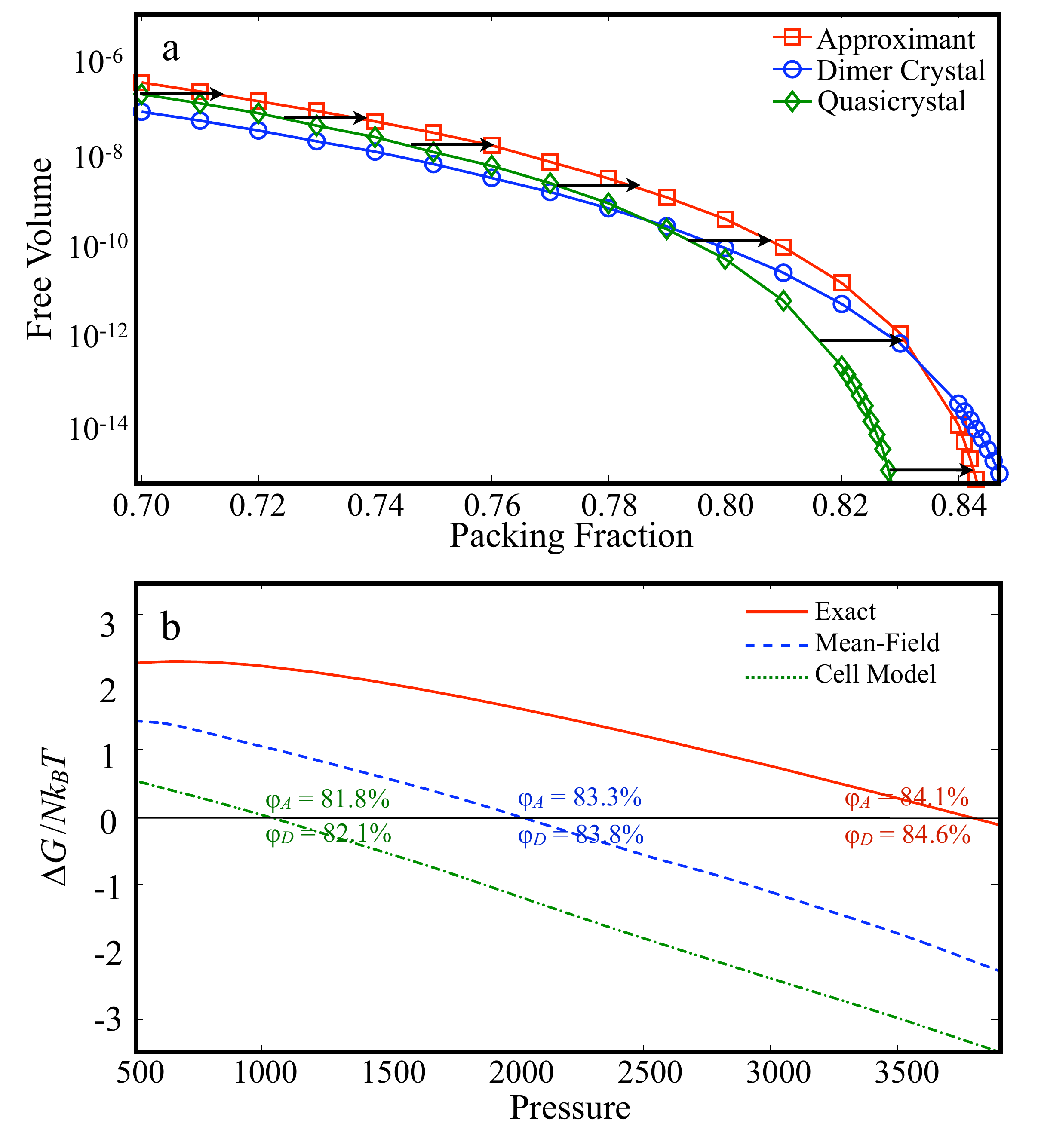}
\caption{Relative stability of the dimer crystal, quasicrystal, and quasicrystal approximant. (a) Up to packing density $83\%$ the dimer crystal has lower average free volume per particle. This helps to stabilize the approximant entropically. At high packing densities the dimer crystal should eventually have the highest average free volume, because its maximally achievable density is the highest of the three candidate structures. (b) Comparison of the Gibbs free energy differences between the dimer crystal and the approximant using the exact Frenkel-Ladd method, the mean-field approximation, and the cell-model approximations. The transition is predicted with all three methods even though the critical densities $\phi_A$ (approximant) and $\phi_D$ (dimer crystal) vary slightly. \label{fig:FreeVolume}}
\end{figure}

We calculate the mean-log average of the free volumes $v_{f,\text{ML}}$ of tetrahedra (Eq.~\ref{eq:vfML}) in the approximant, the dimer crystal, and the quasicrystal using the shooting method described in Section~\ref{subsubsection:shooting}. The results are presented in Fig.~\ref{fig:FreeVolume}a. Whereas particles in the quasicrystal generally have a smaller $v_{f,\text{ML}}$ than in the approximant, the curves are shifted along the abscissa relative to one another by a fixed amount as indicated with arrows in Fig.~\ref{fig:FreeVolume}a. This implies an identical thermodynamics for the quasicrystal and the approximant except for their different maximum packing densities. Indeed, tetrahedra experience similar local environments in the quasicrystal and its approximant.

In contrast, the mean log free volume of the dimer crystal decays much more slowly with packing density and intersects the two other free volume curves. This finding suggests that the approximant relaxes more efficiently during expansion, creating free volume for the particles more readily. Note that the packing density where the two curves cross is considerably below $84\%$, the density where the approximant becomes thermodynamically unstable, which underscores the significance of collective motions of particles in stabilizing the approximant even at very high densities. 

The importance of collective motions may be further inferred by comparing the free energy difference estimated from a mean-field approximation with the exact value. As shown in Fig.~\ref{fig:FreeVolume}b, the mean-field approximation underestimates the stability of the approximant, which indicates that entropic contributions from collective motions are significant. We suspect that slight rearrangements of particles in the approximant during expansion also increase its stability at lower packing densities. This is confirmed by estimating $\Delta{G}$ from a cell model approximation. The cell model is similar to the mean-field approximation except that free volumes are calculated for a non-equilibrated structure obtained by isotropically expanding the densest packing to a given packing density~\cite{VegaMonsonMolPhys1992}. As shown in Fig.~\ref{fig:FreeVolume}b, the cell-model approximation underestimates the stability of the approximant even more than the mean-field approximation, which suggests the significance of small local rearrangements that occur while the structure is equilibrated after expansion. 

\subsubsection{Dynamics in the approximant}

Correlated motions of tetrahedra are observed in long simulations of both the approximant and the quasicrystal at all densities. These motions are most apparent at packing densities below $65\%$ where they give rise to local structural rearrangements, but they are still present at higher densities in the form of correlated vibrations of clusters of tetrahedra. The fundamental mechanism through which these rearrangements proceed is the rotation of single PDs around their principal axes by multiples of $72^{\circ}$. The rounded, disk-like shape of PDs, compared to tetrahedra with their sharp corners allows an easy rotation even in relatively dense configurations.

\begin{figure}
\includegraphics[width=\columnwidth]{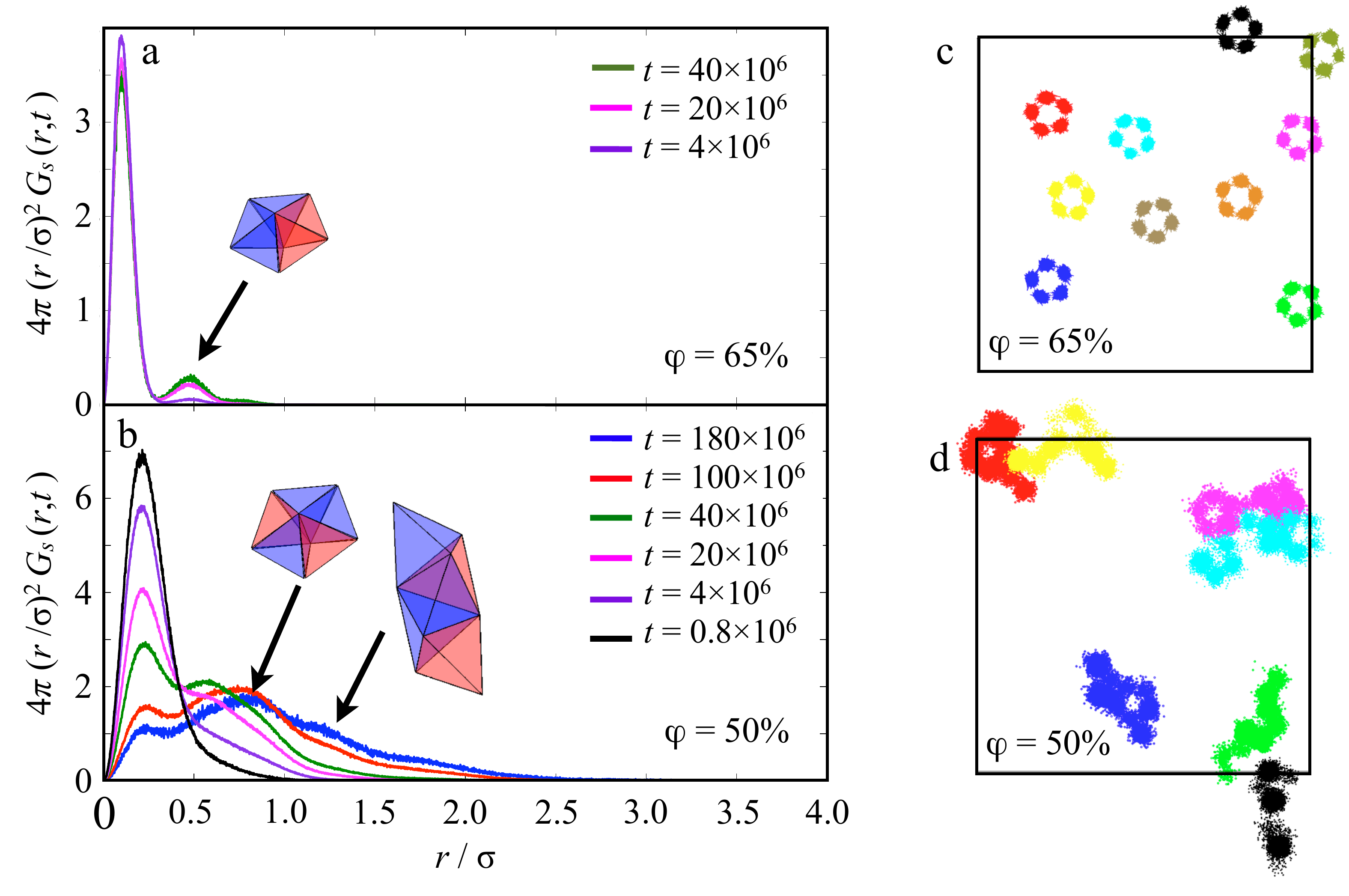}
\caption{Particle dynamics in the quasicrystal approximant. (a,b) The self-parts of the van Hove correlation functions at $\phi=65\%$ (a) and $\phi=50\%$ (b) show various peaks, which indicates that the particles do not move continuously but have to overcome (geometric) barriers. The peak positions correspond to different levels of nearest neighbor distances in the underlying PD network. (c,d) The trajectories of particles with the highest mobility are plotted. At high density, $\phi=65\%$ (c), tetrahedra move along the edges of pentagons. This motion corresponds to rotations of the PDs in log centers. At intermediate densities, $\phi=50\%$ (d), neighboring PDs start to rotate and the tetrahedra are more mobile. In the infinite time limit the tetrahedra can diffuse through successive PD rotations. \label{fig:Dynamics}}
\end{figure}

The rotation of PDs is confirmed by observing several peaks in $G_s(r,t)$, the self-part of the van Hove correlation function~\cite{vanHovePR1954}, which implies that the tetrahedra indeed move between discrete sites separated by geometric barriers (Fig.~\ref{fig:Dynamics}a,b). As reported in our earlier work~\cite{HajiAkbariEtAl2009}, each tetrahedron in the quasicrystal and the approximant is part of a spanning network of interpenetrating PDs (that is, PDs that share a tetrahedron). The locations of the peaks in $G_s(r,t)$ correspond to the characteristic distances of the nearest neighbor distances in the spanning network. 

\begin{figure}
\includegraphics[width=\columnwidth]{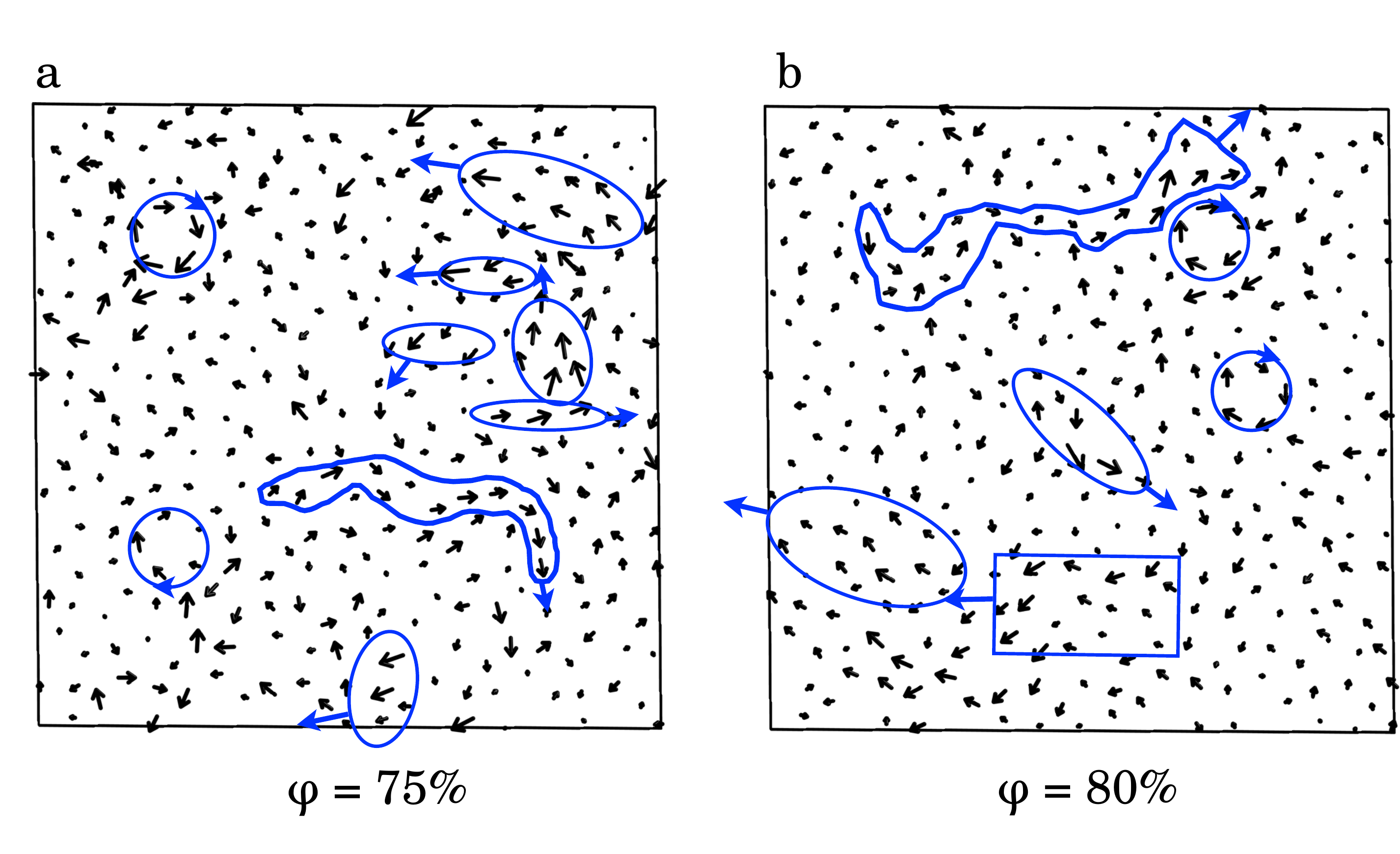}
\caption{
Correlated motion of clusters in a slab of the approximant at (a) $\phi=75\%$ and (b) $\phi=80\%$. Dark arrows correspond to the direction towards which each particle has moved after $t=5\times10^7$ Monte Carlo cycles; the length of each arrow is twice the distance the corresponding particle had travelled. There are several clusters of neighboring tetrahedra moving collectively. A few of these clusters are highlighted in blue. Not surprisingly, the mobility is higher at $\phi=75\%$ as evidenced by longer arrows.\label{fig:vibHighDensity}}
\end{figure}

We observe that not all PDs are equally likely to rotate. At high densities, the PDs capping the 12-fold rings in the center of logs (shown in green in Fig. 1b)\cite{HajiAkbariEtAl2009} rotate more frequently as they are spatially separated from the rest of the structure. This can be seen in the trajectories of the high-mobility particles in the approximant at $\phi=65\%$ (Fig.~\ref{fig:Dynamics}c). Close to melting, however, rotations involve the full network of neighboring PDs, which allows the particles to diffuse over arbitrary distances (Fig.~\ref{fig:Dynamics}d). The underlying dynamics is identical in the quasicrystal. However the presence of defects leads to higher mobility in the quasicrystals that form in simulation as compared with "perfect" quasicrystals. Both the quasicrystal and the approximant exhibit some 'liquid-like' behavior since unlike simple crystals, diffusion can take place in these systems even in the absence of defects.

At packing densities beyond $65\%$, PD rotations become extremely unlikely, but clusters of tetrahedra, including PDs, can still vibrate collectively. Figs.~\ref{fig:vibHighDensity}a-b show such correlated motions  occurring in a time period of $50$ million Monte Carlo cycles in a layer of the approximant at $\phi=75\%$ and $\phi=80\%$ respectively. The vibrations are extremely slow, but their existence adds additional entropy to the system making the mean-field approximation and the cell model inaccurate. No dynamics is observed in the dimer crystal.

In general, thermodynamically equivalent local rearrangements are a characteristic feature of quasicrystals and their approximants. The transformation among these takes place via phason modes~\cite{BakPRB1985, LubenskyPRB1985, LevinePRB1986}. Elementary excitations are phason flips, which previously have been observed with high-resolution transmission electron microscopy~\cite{EdagawaPRL2000} and in simulations of two dimensional model systems~\cite{EngelPRB2010}.

\section{Discussion and conclusion\label{section:conclusion}}
In general one might expect a `simple' structure like the dimer crystal to form more easily than `complex' structures like the quasicrystal or its approximant. The observation that tetrahedra defy this expectation suggests that structural complexity is not always a good indicator of thermodynamic stability. Indeed, although it has been argued in the literature~\cite{TJPRE2010} that the dimer crystal first reported by~\cite{ChenEtAlarxiv2010,Chen2010} and studied here might be the stable phase even at densities where the quasicrystal is reproducibly observed (down to densities of $50\%$), our free energy calculations demonstrate that the dimer crystal is in fact preferred thermodynamically only at very high densities (above $84\%$).  On the other hand, insofar as structural complexity increases a system's entropy, structurally complex arrangements of hard particles may be thermodynamically preferred over simpler ones.

Indeed, we have shown that the structural features of the quasicrystal and the approximant allow for more complex dynamics than the dimer crystal at moderate and high densities as manifested in the behavior of the free volume as a function of packing density and the collective motions in the form of PD rotations. The existence of the PD network facilitates collective particle motions at low densities. Although rearrangements become vanishingly unlikely at higher densities, they appear to contribute additional entropy to the system and stabilize it over the dimer crystal, in which each particle can only `rattle' independently in its own cage. Rearrangements are impossible in the dimer crystal because no rearrangeable network exists there. 

The superior stability of the quasicrystal and its approximant relative to the dimer crystal may also be attributed to the presence of almost-perfect face-to-face contacts between tetrahedra. There is a natural tendency for hard polyhedra to optimize face-to-face contacts at high densities in order to maximize configurational entropy. For instance, there are an infinite number of cubic arrangements of hard cubes with packing fraction one, but among them the simple cubic lattice, where all cubes are perfectly face-to-face, has the highest entropy and is thermodynamically stable~\cite{GrothJCP2001}.

Within the approximant, we observe that face-to-face contacts between neighboring tetrahedra are nearly perfect in the sense that the touching faces are not significantly shifted with respect to one another. This is not true in the dimer crystal where inter-dimer face-to-face contacts are shifted and therefore not close to being perfect. Abundance of strong face-to-face contacts makes the PD network more rearrangeable and collective motions of particles more feasible, which in turn leads to a higher entropy and superior stability.

\begin{figure}
\includegraphics[width=\columnwidth]{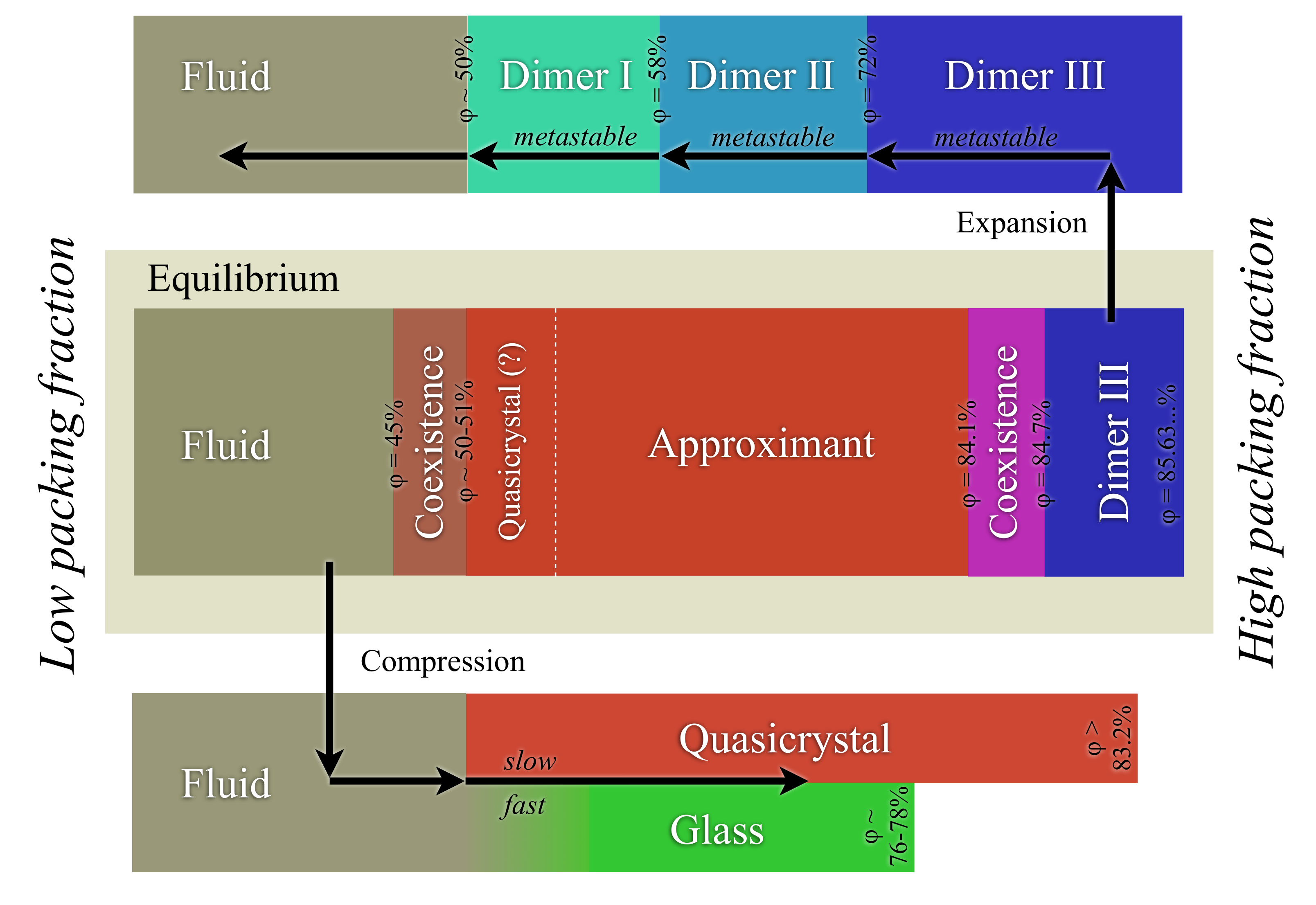}
\caption{Schematic phase diagram of hard tetrahedra summarizing our findings. In thermodynamic equilibrium the Dimer III crystal and the approximant are stable (Middle panel). In compression simulations the approximant is never observed, and only the quasicrystal forms. If crystallization is suppressed, then a jammed packing with local tetrahedral order forms~\cite{HajiAkbariEtAl2009, JaoshvilliPRL2010} (Lower panel). The transformation of the approximant or quasicrystal directly to and from the Dimer III crystal is not observed in simulation. Instead, during expansion the Dimer III crystal transforms into the Dimer II crystal, and then the Dimer I phase prior to melting to the fluid (Upper Panel).  \label{fig:PhaseDiagram}}
\end{figure}

We summarize our findings in a schematic phase diagram in Fig.~\ref{fig:PhaseDiagram}. We note that hard tetrahedra are one of the few examples of hard particles with two distinct solid phases not mutually related by symmetry breaking. Our results show that entropic effects alone are sufficient for inducing highly nontrivial solid-solid phase transitions. 

Not all phase transformations are accessible in simulations on finite time scales. The observation that simulations only form the quasicrystal but never the approximant suggests that the quasicrystal is kinetically more easily accessible than the approximant -- independent of whether it is thermodynamically preferred or not. This can be attributed to the fact that the transformation of a dodecagonal quasicrystal to one of its approximants proceeds through a process called zipper motion ~\cite{OxborrowPRB1993}, which is extremely slow even in experiment~\cite{TalapinNature2009}.
 Furthermore, transformation to the dimer crystal at packing densities greater than $84\%$ is not observable in simulations, and may be unobservable in experiments, due to the extremely slow kinetics at such high densities. 

In conclusion, we have shown that the quasicrystal and its approximant are thermodynamically favored over the dimer crystal at all experimentally realizable packing densities. We also observe a very rich dynamical behavior in the quasicrystal and its approximant induced by rotations of pentagonal dipyramids within an interconnected network. We have shown the significance of collective motions in stabilizing the approximant for a wide range of packing densities.

\section{Acknowledgments}
The authors gratefully acknowledge discussions with D. Frenkel, D. Kofke, R. Petschek and A. Schultz regarding free energy calculations, with P. Charbonneau, I. Nezbeda, and C. Vega concerning free volume calculations, with P. Palffy-Muhoray regarding dynamics, and with V. Elser and Y. Kallus regarding the dimer crystal. This work was supported in part by the U. S. Air Force Office of Scientific Research (FA9550-06-1-0337) and by a U.S. Department of Defense National Security Science and Engineering Faculty Fellowship (N00244-09-1-0062). M.E. acknowledges support from the Deutsche Forschungsgemeinschaft.  A.H-A acknowledges support from the University of Michigan Rackham Predoctoral Fellowship program.

\bibliographystyle{apsrev}
\bibliography{HardTetBib}

\begin{thebibliography}{58}
\expandafter\ifx\csname natexlab\endcsname\relax\def\natexlab#1{#1}\fi
\expandafter\ifx\csname bibnamefont\endcsname\relax
  \def\bibnamefont#1{#1}\fi
\expandafter\ifx\csname bibfnamefont\endcsname\relax
  \def\bibfnamefont#1{#1}\fi
\expandafter\ifx\csname citenamefont\endcsname\relax
  \def\citenamefont#1{#1}\fi
\expandafter\ifx\csname url\endcsname\relax
  \def\url#1{\texttt{#1}}\fi
\expandafter\ifx\csname urlprefix\endcsname\relax\def\urlprefix{URL }\fi
\providecommand{\bibinfo}[2]{#2}
\providecommand{\eprint}[2][]{\url{#2}}

\bibitem[{\citenamefont{Glotzer and
  Solomon}(2007)}]{GlotzerSolomonNatureMatrial2007}
\bibinfo{author}{\bibfnamefont{S.~C.} \bibnamefont{Glotzer}} \bibnamefont{and}
  \bibinfo{author}{\bibfnamefont{M.}~\bibnamefont{Solomon}},
  \bibinfo{journal}{Nat. Mater.} \textbf{\bibinfo{volume}{6}},
  \bibinfo{pages}{567} (\bibinfo{year}{2007}).

\bibitem[{\citenamefont{El-Sayed}(2004)}]{ElSayedAccChemRes2004}
\bibinfo{author}{\bibfnamefont{M.~A.} \bibnamefont{El-Sayed}},
  \bibinfo{journal}{Acc. Chem. Res.} \textbf{\bibinfo{volume}{37}},
  \bibinfo{pages}{326} (\bibinfo{year}{2004}).

\bibitem[{\citenamefont{Burda et~al.}(2005)\citenamefont{Burda, Chen,
  Narayanan, and El-Sayed}}]{BurdaChemRev2005}
\bibinfo{author}{\bibfnamefont{C.}~\bibnamefont{Burda}},
  \bibinfo{author}{\bibfnamefont{X.}~\bibnamefont{Chen}},
  \bibinfo{author}{\bibfnamefont{R.}~\bibnamefont{Narayanan}},
  \bibnamefont{and} \bibinfo{author}{\bibfnamefont{M.~A.}
  \bibnamefont{El-Sayed}}, \bibinfo{journal}{Chem. Rev.}
  \textbf{\bibinfo{volume}{105}}, \bibinfo{pages}{1025} (\bibinfo{year}{2005}).

\bibitem[{\citenamefont{Murphy et~al.}(2005)\citenamefont{Murphy, Sau, Gole,
  Orendorff, Gao, Gou, Hunyadi, and Li}}]{MurphyPhysChemB2005}
\bibinfo{author}{\bibfnamefont{C.~J.} \bibnamefont{Murphy}},
  \bibinfo{author}{\bibfnamefont{T.~K.} \bibnamefont{Sau}},
  \bibinfo{author}{\bibfnamefont{A.~M.} \bibnamefont{Gole}},
  \bibinfo{author}{\bibfnamefont{C.~J.} \bibnamefont{Orendorff}},
  \bibinfo{author}{\bibfnamefont{J.}~\bibnamefont{Gao}},
  \bibinfo{author}{\bibfnamefont{L.}~\bibnamefont{Gou}},
  \bibinfo{author}{\bibfnamefont{S.~E.} \bibnamefont{Hunyadi}},
  \bibnamefont{and} \bibinfo{author}{\bibfnamefont{T.}~\bibnamefont{Li}},
  \bibinfo{journal}{Phys. Chem. B} \textbf{\bibinfo{volume}{109}},
  \bibinfo{pages}{13857} (\bibinfo{year}{2005}).

\bibitem[{\citenamefont{Nie et~al.}(2010)\citenamefont{Nie, Petukhova, and
  Kumacheva}}]{NieNatureNano2010}
\bibinfo{author}{\bibfnamefont{Z.}~\bibnamefont{Nie}},
  \bibinfo{author}{\bibfnamefont{A.}~\bibnamefont{Petukhova}},
  \bibnamefont{and}
  \bibinfo{author}{\bibfnamefont{E.}~\bibnamefont{Kumacheva}},
  \bibinfo{journal}{Nature Nano.} \textbf{\bibinfo{volume}{5}},
  \bibinfo{pages}{15} (\bibinfo{year}{2010}).

\bibitem[{\citenamefont{Podsiadlo et~al.}(2011)\citenamefont{Podsiadlo,
  Krylova, Demortiere, and Shevchenko}}]{Shevchenko2011}
\bibinfo{author}{\bibfnamefont{P.}~\bibnamefont{Podsiadlo}},
  \bibinfo{author}{\bibfnamefont{G.~V.} \bibnamefont{Krylova}},
  \bibinfo{author}{\bibfnamefont{A.}~\bibnamefont{Demortiere}},
  \bibnamefont{and} \bibinfo{author}{\bibfnamefont{E.~V.}
  \bibnamefont{Shevchenko}}, \bibinfo{journal}{J. Nanopart. Res.}
  \textbf{\bibinfo{volume}{13}}, \bibinfo{pages}{15} (\bibinfo{year}{2011}).

\bibitem[{\citenamefont{Kim et~al.}(2004)\citenamefont{Kim, Connor, Song,
  Kuykendall, and Yang}}]{KimAngewChemIntEd2004}
\bibinfo{author}{\bibfnamefont{F.}~\bibnamefont{Kim}},
  \bibinfo{author}{\bibfnamefont{S.}~\bibnamefont{Connor}},
  \bibinfo{author}{\bibfnamefont{H.}~\bibnamefont{Song}},
  \bibinfo{author}{\bibfnamefont{T.}~\bibnamefont{Kuykendall}},
  \bibnamefont{and} \bibinfo{author}{\bibfnamefont{P.}~\bibnamefont{Yang}},
  \bibinfo{journal}{Angew. Chem. Int. Ed.} \textbf{\bibinfo{volume}{43}},
  \bibinfo{pages}{3673} (\bibinfo{year}{2004}).

\bibitem[{\citenamefont{Demortiere et~al.}(2008)\citenamefont{Demortiere,
  Launois, Gaubet, Albouy, and Petit}}]{DemortiereJPhysChemB2008}
\bibinfo{author}{\bibfnamefont{A.}~\bibnamefont{Demortiere}},
  \bibinfo{author}{\bibfnamefont{P.}~\bibnamefont{Launois}},
  \bibinfo{author}{\bibfnamefont{N.}~\bibnamefont{Gaubet}},
  \bibinfo{author}{\bibfnamefont{P.-A.} \bibnamefont{Albouy}},
  \bibnamefont{and} \bibinfo{author}{\bibfnamefont{C.}~\bibnamefont{Petit}},
  \bibinfo{journal}{J. Phys. Chem. B} \textbf{\bibinfo{volume}{112}},
  \bibinfo{pages}{14583} (\bibinfo{year}{2008}).

\bibitem[{\citenamefont{Berenschot et~al.}(2009)\citenamefont{Berenschot, Tas,
  Jansen, and Elwenspoek}}]{Berenschot2009}
\bibinfo{author}{\bibfnamefont{J.~W.} \bibnamefont{Berenschot}},
  \bibinfo{author}{\bibfnamefont{N.~R.} \bibnamefont{Tas}},
  \bibinfo{author}{\bibfnamefont{H.~V.} \bibnamefont{Jansen}},
  \bibnamefont{and}
  \bibinfo{author}{\bibfnamefont{M.}~\bibnamefont{Elwenspoek}},
  \bibinfo{journal}{Nanotech} \textbf{\bibinfo{volume}{20}},
  \bibinfo{pages}{475302} (\bibinfo{year}{2009}).

\bibitem[{\citenamefont{Barrett et~al.}(2009)\citenamefont{Barrett, Dickinson,
  Ahmed, Hantschel, Arstila, and Ryan}}]{Barett2009}
\bibinfo{author}{\bibfnamefont{C.~A.} \bibnamefont{Barrett}},
  \bibinfo{author}{\bibfnamefont{C.}~\bibnamefont{Dickinson}},
  \bibinfo{author}{\bibfnamefont{S.}~\bibnamefont{Ahmed}},
  \bibinfo{author}{\bibfnamefont{T.}~\bibnamefont{Hantschel}},
  \bibinfo{author}{\bibfnamefont{K.}~\bibnamefont{Arstila}}, \bibnamefont{and}
  \bibinfo{author}{\bibfnamefont{K.~W.} \bibnamefont{Ryan}},
  \bibinfo{journal}{Nanotech.} \textbf{\bibinfo{volume}{20}},
  \bibinfo{pages}{275605} (\bibinfo{year}{2009}).

\bibitem[{\citenamefont{Manoharan et~al.}(2003)\citenamefont{Manoharan,
  Elsesser, and Pine}}]{ManoharanPineScience2003}
\bibinfo{author}{\bibfnamefont{V.~N.} \bibnamefont{Manoharan}},
  \bibinfo{author}{\bibfnamefont{M.~T.} \bibnamefont{Elsesser}},
  \bibnamefont{and} \bibinfo{author}{\bibfnamefont{D.~J.} \bibnamefont{Pine}},
  \bibinfo{journal}{Science} \textbf{\bibinfo{volume}{301}},
  \bibinfo{pages}{483} (\bibinfo{year}{2003}).

\bibitem[{\citenamefont{Onsager}(1949)}]{Onsager1949}
\bibinfo{author}{\bibfnamefont{L.}~\bibnamefont{Onsager}},
  \bibinfo{journal}{Annals of the New York Academy of Science}
  \textbf{\bibinfo{volume}{51}}, \bibinfo{pages}{627} (\bibinfo{year}{1949}).

\bibitem[{\citenamefont{Kirkwood}(1951)}]{Kirkwood1951}
\bibinfo{author}{\bibfnamefont{J.~E.} \bibnamefont{Kirkwood}}, in
  \emph{\bibinfo{booktitle}{Phase Transformations in Solids}}, edited by
  \bibinfo{editor}{\bibfnamefont{R.}~\bibnamefont{Smoluchowski}},
  \bibinfo{editor}{\bibfnamefont{J.~E.} \bibnamefont{Mayer}}, \bibnamefont{and}
  \bibinfo{editor}{\bibfnamefont{W.~A.} \bibnamefont{Weyl}}
  (\bibinfo{publisher}{Wiley}, \bibinfo{year}{1951}), p.~\bibinfo{pages}{67}.

\bibitem[{\citenamefont{Frenkel}(1999)}]{FrenkelPhysicaA1999}
\bibinfo{author}{\bibfnamefont{D.}~\bibnamefont{Frenkel}},
  \bibinfo{journal}{Physica A} \textbf{\bibinfo{volume}{272}},
  \bibinfo{pages}{376} (\bibinfo{year}{1999}).

\bibitem[{\citenamefont{Bezdek and Kuperberg}(2010)}]{BezdekArxiv2010}
\bibinfo{author}{\bibfnamefont{A.}~\bibnamefont{Bezdek}} \bibnamefont{and}
  \bibinfo{author}{\bibfnamefont{W.}~\bibnamefont{Kuperberg}}
  (\bibinfo{year}{2010}), \bibinfo{note}{arXiv:1008.2398v1}.

\bibitem[{\citenamefont{Mulero}(2008)}]{Nulero2008}
\bibinfo{author}{\bibfnamefont{A.}~\bibnamefont{Mulero}},
  \emph{\bibinfo{title}{Theory and simulation of hard-sphere fluids and related
  systems}} (\bibinfo{publisher}{Springer, Berlin}, \bibinfo{year}{2008}).

\bibitem[{\citenamefont{Eppenga and Frenkel}(1984)}]{EppengaFrenkel1984}
\bibinfo{author}{\bibfnamefont{R.}~\bibnamefont{Eppenga}} \bibnamefont{and}
  \bibinfo{author}{\bibfnamefont{D.}~\bibnamefont{Frenkel}},
  \bibinfo{journal}{Mol. Phys.} \textbf{\bibinfo{volume}{52}},
  \bibinfo{pages}{1303} (\bibinfo{year}{1984}).

\bibitem[{\citenamefont{Veerman and Frenkel}(1990)}]{VeermanPhysRevA1990}
\bibinfo{author}{\bibfnamefont{J.~A.~C.} \bibnamefont{Veerman}}
  \bibnamefont{and} \bibinfo{author}{\bibfnamefont{D.}~\bibnamefont{Frenkel}},
  \bibinfo{journal}{Phys. Rev. A} \textbf{\bibinfo{volume}{41}},
  \bibinfo{pages}{3237} (\bibinfo{year}{1990}).

\bibitem[{\citenamefont{Veerman and Frenkel}(1992)}]{VeermanFrenkel1992}
\bibinfo{author}{\bibfnamefont{J.~A.~C.} \bibnamefont{Veerman}}
  \bibnamefont{and} \bibinfo{author}{\bibfnamefont{D.}~\bibnamefont{Frenkel}},
  \bibinfo{journal}{Phys. Rev. A} \textbf{\bibinfo{volume}{45}},
  \bibinfo{pages}{5632} (\bibinfo{year}{1992}).

\bibitem[{\citenamefont{Vega et~al.}(1992)\citenamefont{Vega, Paras, and
  Monson}}]{VegaJCP1992}
\bibinfo{author}{\bibfnamefont{C.}~\bibnamefont{Vega}},
  \bibinfo{author}{\bibfnamefont{E.~P.~A.} \bibnamefont{Paras}},
  \bibnamefont{and} \bibinfo{author}{\bibfnamefont{P.~A.}
  \bibnamefont{Monson}}, \bibinfo{journal}{J. Chem. Phys.}
  \textbf{\bibinfo{volume}{96}}, \bibinfo{pages}{9060} (\bibinfo{year}{1992}).

\bibitem[{\citenamefont{Bolhuis and Frenkel}(1997)}]{BolhuisFrenkel1997}
\bibinfo{author}{\bibfnamefont{P.}~\bibnamefont{Bolhuis}} \bibnamefont{and}
  \bibinfo{author}{\bibfnamefont{D.}~\bibnamefont{Frenkel}},
  \bibinfo{journal}{J. Chem. Phys.} \textbf{\bibinfo{volume}{106}},
  \bibinfo{pages}{666} (\bibinfo{year}{1997}).

\bibitem[{\citenamefont{Camp and Allen}(1997)}]{CampAllen1997}
\bibinfo{author}{\bibfnamefont{P.~J.} \bibnamefont{Camp}} \bibnamefont{and}
  \bibinfo{author}{\bibfnamefont{M.~P.} \bibnamefont{Allen}},
  \bibinfo{journal}{J. Chem. Phys.} \textbf{\bibinfo{volume}{106}},
  \bibinfo{pages}{6681} (\bibinfo{year}{1997}).

\bibitem[{\citenamefont{John et~al.}(2008)\citenamefont{John, Juhlin, and
  Escobedo}}]{JohnEscobedo2008}
\bibinfo{author}{\bibfnamefont{B.~S.} \bibnamefont{John}},
  \bibinfo{author}{\bibfnamefont{C.}~\bibnamefont{Juhlin}}, \bibnamefont{and}
  \bibinfo{author}{\bibfnamefont{F.~A.} \bibnamefont{Escobedo}},
  \bibinfo{journal}{J. Chem. Phys.} \textbf{\bibinfo{volume}{128}},
  \bibinfo{pages}{044909} (\bibinfo{year}{2008}).

\bibitem[{\citenamefont{Radu et~al.}(2009)\citenamefont{Radu, Pfleiderer, and
  Schilling}}]{RaduSchilling2009}
\bibinfo{author}{\bibfnamefont{M.}~\bibnamefont{Radu}},
  \bibinfo{author}{\bibfnamefont{P.}~\bibnamefont{Pfleiderer}},
  \bibnamefont{and}
  \bibinfo{author}{\bibfnamefont{T.}~\bibnamefont{Schilling}},
  \bibinfo{journal}{J. Chem. Phys.} \textbf{\bibinfo{volume}{131}},
  \bibinfo{pages}{164513} (\bibinfo{year}{2009}).

\bibitem[{\citenamefont{Agarwal and
  Escobedo}(2011)}]{EscobedoNatureMaterials2011}
\bibinfo{author}{\bibfnamefont{U.}~\bibnamefont{Agarwal}} \bibnamefont{and}
  \bibinfo{author}{\bibfnamefont{F.~A.} \bibnamefont{Escobedo}},
  \bibinfo{journal}{Nat. Mater.} \textbf{\bibinfo{volume}{10}},
  \bibinfo{pages}{230} (\bibinfo{year}{2011}).

\bibitem[{\citenamefont{Conway and Torquato}(2006)}]{ConwayTorquatoPNAS2006}
\bibinfo{author}{\bibfnamefont{J.~H.} \bibnamefont{Conway}} \bibnamefont{and}
  \bibinfo{author}{\bibfnamefont{S.}~\bibnamefont{Torquato}},
  \bibinfo{journal}{Proc. Natl. Acad. Sci. USA} \textbf{\bibinfo{volume}{103}},
  \bibinfo{pages}{10612} (\bibinfo{year}{2006}).

\bibitem[{\citenamefont{Chen}(2008)}]{ChenDSC2008}
\bibinfo{author}{\bibfnamefont{E.~R.} \bibnamefont{Chen}},
  \bibinfo{journal}{Disc. Comp. Geom.} \textbf{\bibinfo{volume}{40}},
  \bibinfo{pages}{214} (\bibinfo{year}{2008}).

\bibitem[{\citenamefont{Torquato and
  Jiao}(2009{\natexlab{a}})}]{TorquatoJiaoNature2009}
\bibinfo{author}{\bibfnamefont{S.}~\bibnamefont{Torquato}} \bibnamefont{and}
  \bibinfo{author}{\bibfnamefont{Y.}~\bibnamefont{Jiao}},
  \bibinfo{journal}{Nature} \textbf{\bibinfo{volume}{460}},
  \bibinfo{pages}{876} (\bibinfo{year}{2009}{\natexlab{a}}).

\bibitem[{\citenamefont{Haji-Akbari et~al.}(2009)\citenamefont{Haji-Akbari,
  Engel, Keys, Zheng, Petschek, Palffy-Muhoray, and
  Glotzer}}]{HajiAkbariEtAl2009}
\bibinfo{author}{\bibfnamefont{A.}~\bibnamefont{Haji-Akbari}},
  \bibinfo{author}{\bibfnamefont{M.}~\bibnamefont{Engel}},
  \bibinfo{author}{\bibfnamefont{A.~S.} \bibnamefont{Keys}},
  \bibinfo{author}{\bibfnamefont{X.~Y.} \bibnamefont{Zheng}},
  \bibinfo{author}{\bibfnamefont{R.}~\bibnamefont{Petschek}},
  \bibinfo{author}{\bibfnamefont{P.}~\bibnamefont{Palffy-Muhoray}},
  \bibnamefont{and} \bibinfo{author}{\bibfnamefont{S.~C.}
  \bibnamefont{Glotzer}}, \bibinfo{journal}{Nature}
  \textbf{\bibinfo{volume}{462}}, \bibinfo{pages}{773−}
  (\bibinfo{year}{2009}).

\bibitem[{\citenamefont{Kallus et~al.}(2009)\citenamefont{Kallus, Elser, and
  Gravel}}]{KallusArxiv2009}
\bibinfo{author}{\bibfnamefont{Y.}~\bibnamefont{Kallus}},
  \bibinfo{author}{\bibfnamefont{V.}~\bibnamefont{Elser}}, \bibnamefont{and}
  \bibinfo{author}{\bibfnamefont{S.}~\bibnamefont{Gravel}}
  (\bibinfo{year}{2009}), \bibinfo{note}{arXiv:0910.5226}.

\bibitem[{\citenamefont{Torquato and
  Jiao}(2009{\natexlab{b}})}]{TorquatoJiaoarxiv}
\bibinfo{author}{\bibfnamefont{S.}~\bibnamefont{Torquato}} \bibnamefont{and}
  \bibinfo{author}{\bibfnamefont{Y.}~\bibnamefont{Jiao}}
  (\bibinfo{year}{2009}{\natexlab{b}}), \bibinfo{note}{arXiv:0912.4210}.

\bibitem[{\citenamefont{Chen et~al.}(2010{\natexlab{a}})\citenamefont{Chen,
  Engel, and Glotzer}}]{ChenEtAlarxiv2010}
\bibinfo{author}{\bibfnamefont{E.~R.} \bibnamefont{Chen}},
  \bibinfo{author}{\bibfnamefont{M.}~\bibnamefont{Engel}}, \bibnamefont{and}
  \bibinfo{author}{\bibfnamefont{S.~C.} \bibnamefont{Glotzer}}
  (\bibinfo{year}{2010}{\natexlab{a}}), \bibinfo{note}{arXiv:1001.0586}.

\bibitem[{\citenamefont{Kallus et~al.}(2010)\citenamefont{Kallus, Elser, and
  Gravel}}]{Kallus2010}
\bibinfo{author}{\bibfnamefont{Y.}~\bibnamefont{Kallus}},
  \bibinfo{author}{\bibfnamefont{V.}~\bibnamefont{Elser}}, \bibnamefont{and}
  \bibinfo{author}{\bibfnamefont{S.}~\bibnamefont{Gravel}},
  \bibinfo{journal}{Disc. Comp. Geom.} \textbf{\bibinfo{volume}{44}},
  \bibinfo{pages}{245} (\bibinfo{year}{2010}).

\bibitem[{\citenamefont{Chen et~al.}(2010{\natexlab{b}})\citenamefont{Chen,
  Engel, and Glotzer}}]{Chen2010}
\bibinfo{author}{\bibfnamefont{E.~R.} \bibnamefont{Chen}},
  \bibinfo{author}{\bibfnamefont{M.}~\bibnamefont{Engel}}, \bibnamefont{and}
  \bibinfo{author}{\bibfnamefont{S.~C.} \bibnamefont{Glotzer}},
  \bibinfo{journal}{Disc. Comp. Geom.} \textbf{\bibinfo{volume}{44}},
  \bibinfo{pages}{253} (\bibinfo{year}{2010}{\natexlab{b}}).

\bibitem[{\citenamefont{Torquato and Jiao}(2010)}]{TJPRE2010}
\bibinfo{author}{\bibfnamefont{S.}~\bibnamefont{Torquato}} \bibnamefont{and}
  \bibinfo{author}{\bibfnamefont{Y.}~\bibnamefont{Jiao}},
  \bibinfo{journal}{Phys. Rev. E} \textbf{\bibinfo{volume}{81}},
  \bibinfo{pages}{041310} (\bibinfo{year}{2010}).

\bibitem[{\citenamefont{Jaoshvili et~al.}(2010)\citenamefont{Jaoshvili, Esakia,
  Porrati, and Chaikin}}]{JaoshvilliPRL2010}
\bibinfo{author}{\bibfnamefont{A.}~\bibnamefont{Jaoshvili}},
  \bibinfo{author}{\bibfnamefont{A.}~\bibnamefont{Esakia}},
  \bibinfo{author}{\bibfnamefont{M.}~\bibnamefont{Porrati}}, \bibnamefont{and}
  \bibinfo{author}{\bibfnamefont{P.~M.} \bibnamefont{Chaikin}},
  \bibinfo{journal}{Phys. Rev. Lett.} \textbf{\bibinfo{volume}{104}},
  \bibinfo{pages}{185501} (\bibinfo{year}{2010}).

\bibitem[{\citenamefont{Gibbons}(1970)}]{GibbonsMolPhys1970}
\bibinfo{author}{\bibfnamefont{R.~M.} \bibnamefont{Gibbons}},
  \bibinfo{journal}{Mol. Phys.} \textbf{\bibinfo{volume}{18}},
  \bibinfo{pages}{809} (\bibinfo{year}{1970}).

\bibitem[{\citenamefont{Kolafa and Nezbeda}(1995)}]{KolafaMolPhys1995}
\bibinfo{author}{\bibfnamefont{J.}~\bibnamefont{Kolafa}} \bibnamefont{and}
  \bibinfo{author}{\bibfnamefont{I.}~\bibnamefont{Nezbeda}},
  \bibinfo{journal}{Mol. Phys.} \textbf{\bibinfo{volume}{84}},
  \bibinfo{pages}{421} (\bibinfo{year}{1995}).

\bibitem[{\citenamefont{Frenkel and Smit}(2002)}]{FrenkelSmitBook}
\bibinfo{author}{\bibfnamefont{D.}~\bibnamefont{Frenkel}} \bibnamefont{and}
  \bibinfo{author}{\bibfnamefont{B.}~\bibnamefont{Smit}},
  \emph{\bibinfo{title}{Understanding Molecular Simulation}}
  (\bibinfo{publisher}{Academic Press, San Diego}, \bibinfo{year}{2002}).

\bibitem[{\citenamefont{Frenkel and Ladd}(1984)}]{FrenkelLaddJCP1984}
\bibinfo{author}{\bibfnamefont{D.}~\bibnamefont{Frenkel}} \bibnamefont{and}
  \bibinfo{author}{\bibfnamefont{A.~J.~C.} \bibnamefont{Ladd}},
  \bibinfo{journal}{J. Chem. Phys.} \textbf{\bibinfo{volume}{81}},
  \bibinfo{pages}{3188} (\bibinfo{year}{1984}).

\bibitem[{\citenamefont{Marechal and Dijkstra}(2008)}]{DijkstraPRE2008}
\bibinfo{author}{\bibfnamefont{M.}~\bibnamefont{Marechal}} \bibnamefont{and}
  \bibinfo{author}{\bibfnamefont{M.}~\bibnamefont{Dijkstra}},
  \bibinfo{journal}{Phys. Rev. E} \textbf{\bibinfo{volume}{77}},
  \bibinfo{pages}{061405} (\bibinfo{year}{2008}).

\bibitem[{\citenamefont{Engel}(2011)}]{EngelPRL2011}
\bibinfo{author}{\bibfnamefont{M.}~\bibnamefont{Engel}},
  \bibinfo{journal}{Phys. Rev. Lett.} \textbf{\bibinfo{volume}{108}},
  \bibinfo{pages}{095504} (\bibinfo{year}{2011}).

\bibitem[{\citenamefont{Romano et~al.}(2010)\citenamefont{Romano, Sanz, and
  Sciortino}}]{RomanoJChemPhys2010}
\bibinfo{author}{\bibfnamefont{F.}~\bibnamefont{Romano}},
  \bibinfo{author}{\bibfnamefont{E.}~\bibnamefont{Sanz}}, \bibnamefont{and}
  \bibinfo{author}{\bibfnamefont{F.}~\bibnamefont{Sciortino}},
  \bibinfo{journal}{J. Chem. Phys.} \textbf{\bibinfo{volume}{132}},
  \bibinfo{pages}{184501} (\bibinfo{year}{2010}).

\bibitem[{\citenamefont{Hoover and Ree}(1968)}]{Hoover1968}
\bibinfo{author}{\bibfnamefont{W.~G.} \bibnamefont{Hoover}} \bibnamefont{and}
  \bibinfo{author}{\bibfnamefont{F.~H.} \bibnamefont{Ree}},
  \bibinfo{journal}{J. Chem. Phys.} \textbf{\bibinfo{volume}{49}},
  \bibinfo{pages}{3609} (\bibinfo{year}{1968}).

\bibitem[{\citenamefont{Paras et~al.}(1992)\citenamefont{Paras, Vega, and
  A}}]{VegaMonsonMolPhys1992}
\bibinfo{author}{\bibfnamefont{E.~P.~A.} \bibnamefont{Paras}},
  \bibinfo{author}{\bibfnamefont{C.}~\bibnamefont{Vega}}, \bibnamefont{and}
  \bibinfo{author}{\bibfnamefont{M.~P.} \bibnamefont{A}},
  \bibinfo{journal}{Mol. Phys.} \textbf{\bibinfo{volume}{77}},
  \bibinfo{pages}{803} (\bibinfo{year}{1992}).

\bibitem[{\citenamefont{Stampfii}(1986)}]{StampfiiHelvPhysActa1986}
\bibinfo{author}{\bibfnamefont{P.}~\bibnamefont{Stampfii}},
  \bibinfo{journal}{Helv. Phys. Acta} \textbf{\bibinfo{volume}{59}},
  \bibinfo{pages}{1260} (\bibinfo{year}{1986}).

\bibitem[{\citenamefont{Keys et~al.}(2011)\citenamefont{Keys, Iacovella, and
  Glotzer}}]{KeysAnnRevCondMatPhys2011}
\bibinfo{author}{\bibfnamefont{A.~S.} \bibnamefont{Keys}},
  \bibinfo{author}{\bibfnamefont{C.~R.} \bibnamefont{Iacovella}},
  \bibnamefont{and} \bibinfo{author}{\bibfnamefont{S.~C.}
  \bibnamefont{Glotzer}}, \bibinfo{journal}{Ann. Rev. Cond. Mat. Phys.}
  \textbf{\bibinfo{volume}{2}}, \bibinfo{pages}{263} (\bibinfo{year}{2011}).

\bibitem[{\citenamefont{Harismiadis et~al.}(1996)\citenamefont{Harismiadis,
  Vorholz, and Panagiotopoulos}}]{HarismiadisJChemPhys1996}
\bibinfo{author}{\bibfnamefont{V.~I.} \bibnamefont{Harismiadis}},
  \bibinfo{author}{\bibfnamefont{J.}~\bibnamefont{Vorholz}}, \bibnamefont{and}
  \bibinfo{author}{\bibfnamefont{A.~Z.} \bibnamefont{Panagiotopoulos}},
  \bibinfo{journal}{J. Chem. Phys.} \textbf{\bibinfo{volume}{105}},
  \bibinfo{pages}{8469} (\bibinfo{year}{1996}).

\bibitem[{\citenamefont{van Hove}(1954)}]{vanHovePR1954}
\bibinfo{author}{\bibfnamefont{L.}~\bibnamefont{van Hove}},
  \bibinfo{journal}{Phys. Rev.} \textbf{\bibinfo{volume}{95}},
  \bibinfo{pages}{249} (\bibinfo{year}{1954}).

\bibitem[{\citenamefont{Kob et~al.}(1997)\citenamefont{Kob, Donati, Plimpton,
  Poole, and Glotzer}}]{KobPRL1997}
\bibinfo{author}{\bibfnamefont{W.}~\bibnamefont{Kob}},
  \bibinfo{author}{\bibfnamefont{C.}~\bibnamefont{Donati}},
  \bibinfo{author}{\bibfnamefont{S.~J.} \bibnamefont{Plimpton}},
  \bibinfo{author}{\bibfnamefont{P.~H.} \bibnamefont{Poole}}, \bibnamefont{and}
  \bibinfo{author}{\bibfnamefont{S.~C.} \bibnamefont{Glotzer}},
  \bibinfo{journal}{Phys. Rev. Lett.} \textbf{\bibinfo{volume}{79}},
  \bibinfo{pages}{2827} (\bibinfo{year}{1997}).

\bibitem[{\citenamefont{Bak}(1985)}]{BakPRB1985}
\bibinfo{author}{\bibfnamefont{P.}~\bibnamefont{Bak}}, \bibinfo{journal}{Phys.
  Rev. B} \textbf{\bibinfo{volume}{32}}, \bibinfo{pages}{5764}
  (\bibinfo{year}{1985}).

\bibitem[{\citenamefont{Lubensky and Ramaswamy}(1985)}]{LubenskyPRB1985}
\bibinfo{author}{\bibfnamefont{T.~C.} \bibnamefont{Lubensky}} \bibnamefont{and}
  \bibinfo{author}{\bibfnamefont{S.}~\bibnamefont{Ramaswamy}},
  \bibinfo{journal}{Phys. Rev. B} \textbf{\bibinfo{volume}{32}},
  \bibinfo{pages}{7444} (\bibinfo{year}{1985}).

\bibitem[{\citenamefont{Levine and Steinhardt}(1986)}]{LevinePRB1986}
\bibinfo{author}{\bibfnamefont{D.}~\bibnamefont{Levine}} \bibnamefont{and}
  \bibinfo{author}{\bibfnamefont{P.~J.} \bibnamefont{Steinhardt}},
  \bibinfo{journal}{Phys. Rev. B} \textbf{\bibinfo{volume}{34}},
  \bibinfo{pages}{596} (\bibinfo{year}{1986}).

\bibitem[{\citenamefont{Edagawa et~al.}(2000)\citenamefont{Edagawa, Suzuki, and
  Takeuchi}}]{EdagawaPRL2000}
\bibinfo{author}{\bibfnamefont{K.}~\bibnamefont{Edagawa}},
  \bibinfo{author}{\bibfnamefont{K.}~\bibnamefont{Suzuki}}, \bibnamefont{and}
  \bibinfo{author}{\bibfnamefont{S.}~\bibnamefont{Takeuchi}},
  \bibinfo{journal}{Phys. Rev. Lett.} \textbf{\bibinfo{volume}{85}},
  \bibinfo{pages}{1674} (\bibinfo{year}{2000}).

\bibitem[{\citenamefont{Engel et~al.}(2010)\citenamefont{Engel, Umezaki,
  Trebin, and Odagaki}}]{EngelPRB2010}
\bibinfo{author}{\bibfnamefont{M.}~\bibnamefont{Engel}},
  \bibinfo{author}{\bibfnamefont{M.}~\bibnamefont{Umezaki}},
  \bibinfo{author}{\bibfnamefont{H.-R.} \bibnamefont{Trebin}},
  \bibnamefont{and} \bibinfo{author}{\bibfnamefont{T.}~\bibnamefont{Odagaki}},
  \bibinfo{journal}{Phys. Rev. B} \textbf{\bibinfo{volume}{82}},
  \bibinfo{pages}{134206} (\bibinfo{year}{2010}).

\bibitem[{\citenamefont{Groh and Mulder}(2001)}]{GrothJCP2001}
\bibinfo{author}{\bibfnamefont{B.}~\bibnamefont{Groh}} \bibnamefont{and}
  \bibinfo{author}{\bibfnamefont{B.}~\bibnamefont{Mulder}},
  \bibinfo{journal}{J. Chem. Phys.} \textbf{\bibinfo{volume}{114}},
  \bibinfo{pages}{3653} (\bibinfo{year}{2001}).

\bibitem[{\citenamefont{Oxborrow and Henley}(1993)}]{OxborrowPRB1993}
\bibinfo{author}{\bibfnamefont{M.}~\bibnamefont{Oxborrow}} \bibnamefont{and}
  \bibinfo{author}{\bibfnamefont{C.~L.} \bibnamefont{Henley}},
  \bibinfo{journal}{Phys. Rev. B.} \textbf{\bibinfo{volume}{48}},
  \bibinfo{pages}{6966} (\bibinfo{year}{1993}).

\bibitem[{\citenamefont{Talapin et~al.}(2009)\citenamefont{Talapin, Shevchenko,
  Bodnarchuk, Ye, Chen, and Murray}}]{TalapinNature2009}
\bibinfo{author}{\bibfnamefont{D.~V.} \bibnamefont{Talapin}},
  \bibinfo{author}{\bibfnamefont{E.~V.} \bibnamefont{Shevchenko}},
  \bibinfo{author}{\bibfnamefont{M.~I.} \bibnamefont{Bodnarchuk}},
  \bibinfo{author}{\bibfnamefont{X.}~\bibnamefont{Ye}},
  \bibinfo{author}{\bibfnamefont{J.}~\bibnamefont{Chen}}, \bibnamefont{and}
  \bibinfo{author}{\bibfnamefont{C.~B.} \bibnamefont{Murray}},
  \bibinfo{journal}{Nature} \textbf{\bibinfo{volume}{461}},
  \bibinfo{pages}{964} (\bibinfo{year}{2009}).

\end{thebibliography}

\end{document}